\documentclass{article}
\usepackage{graphicx}  
\usepackage{amsmath}   
\usepackage{amssymb}   
\usepackage{bm} 
\usepackage{dcolumn}
\usepackage{color}
\usepackage{mathrsfs}
\usepackage{amsfonts}
\usepackage{varioref}
\RequirePackage[colorlinks,citecolor=blue,urlcolor=magenta,linkcolor=blue]{hyperref}
\def\be{\begin{equation}}
\def\ee{\end{equation}}

\labelformat{section}{Section #1} 
\labelformat{subsection}{Section #1} 
\labelformat{subsubsection}{Section #1}
\labelformat{subsubsubsection}{Section #1}
\labelformat{equation}{Eq.~(#1)} 
\labelformat{figure}{Fig.~#1} 
\labelformat{subfigure}{Fig.~\thefigure#1} 
\labelformat{table}{Tab.~#1} 
\labelformat{appendix}{Appendix #1}

\title{\bf Vacuum polarisation of Dirac fermions in the cosmological  de Sitter-global monopole spacetime}
\author{Md Sabir Ali$^1$\footnote{alimd.sabir3@gmail.com},\,\,\, Sourav Bhattacharya$^2$\footnote{sbhatta@iitrpr.ac.in}\\
$^{1,2}$\small{Department of Physics, Indian Institute of Technology Ropar, Rupnagar, Punjab 140 001, India}\\
$^{1}$\small{Department of Physical Sciences, Indian Institute of Science Education and Research Kolkata,}\\
\small Nadia, West Bengal 741 246, India\footnote{Current affiliation.}}
\begin{document}

\maketitle
\begin{abstract}
\noindent
We study the vacuum polarisation effects of the Dirac fermionic field induced by a pointlike global monopole located in the cosmological de Sitter spacetime. First we derive the four orthonormal Dirac modes in this background in a closed form. Quantising the field using these modes, we then compute the fermionic condensate, $\langle 0| \overline{\Psi} \Psi | 0\rangle$, as well as the  vacuum expectation value of the energy-momentum tensor for a massive Dirac  field, regularised in a particular way. We have used  the Abel-Plana summation formula in order to extract the  global monopole contribution  to these quantities and have investigated their variations numerically with respect to relevant parameters.  
\end{abstract}

{\bf Keywords :} de Sitter spacetime, topological defects, fermions, vacuum polarisation 

\newpage

\tableofcontents

\section{Introduction}

The early inflationary period of our universe is a phase of very rapid and nearly exponential  accelerated expansion. This phase  was proposed in order to resolve the three  puzzles of the standard  Big Bang cosmology -- the horizon problem, the flatness problem and the problem of the hitherto unobserved magnetic monopoles~\cite{Linde, Weinberg:2008zzc} (also references therein). The inflationary paradigm also satisfactorily explains how the primordial perturbations, which were quantum field theoretic in nature initially, grew, became large and classical and eventually developed into the large scale cosmic structures as we observe them today.  After the end of the inflation, our universe entered respectively the phases of radiation and matter domination. These phases were also expanding but not with acceleration.  Interestingly,  the astrophysical  data from  high redshift supernovae, the galaxy cluster and the cosmic microwave background suggest that our current universe is also undergoing a phase of accelerated expansion (see \cite{Riess:2006fw} and references therein). In order to drive such accelerated expansions, the universe should be endowed with some exotic matter with negative isotropic pressure, called the dark energy. The simplest and phenomenologically one of the most successful model of the dark energy is simply the positive cosmological constant, and the corresponding solution of the Einstein equation is known as the de Sitter spacetime, having an exponential scale factor and hence a  constant Hubble rate.   

Quantum field theory in the de Sitter background might thus make interesting physical predictions  whose imprint can be found  for example, in the cosmic microwave background. A very important topic in this area is the non-perturbative infrared or secular effect at late times, which is likely to have connection to the cosmological constant and the cosmic coincidence problem, e.g.~\cite{Woodard:2014jba} and references therein. 

In this paper, we shall however be interested in the field theoretic effects in the de Sitter universe endowed with topological defects. Such defects might have  created via some symmetry breaking phase transitions in the early universe after the Big Bang. They may be in the form of cosmic strings, the global monopoles, domain walls and textures~\cite{Vile82, Vilenkin:2000jqa, Kibble:1976sj, Kibble}. Our interest in this paper will be the  global monopoles, first proposed in~\cite{Starobinsky} (see also~\cite{Barriola:1989hx, Ola}). It is a spherically symmetric topologically stable gravitational defect with a deficit in the solid angle $4\pi$.  This  may result from a global-symmetry breaking ($O(3) \to U(1)$) of a self interacting scalar triplet.  Such defects might have singular structure at the centre (like the pointlike global monopole). However, such singularity can be relaxed by taking a finite core and allowing the core to inflate~\cite{Cho}.  The topological defect that has perhaps received more attention than the others is the cosmic string~\cite{Vilenkin:2000jqa}, and references therein. It is a cylindrically symmetric spacetime with a $\delta$-function like line singularity along the symmetry axis. Note that although such defects introduce inhomogeneity, their role in the primordial structure formation is highly suppressed~\cite{Perivolaropoulos:2005wa}. Nevertheless, such defects may also create other physical effects like the emission of gravitational waves and lensing~\cite{Damour:2000wa, Sazhin:2006kf}. In particular, for a realistic situation like a network of cosmic strings, the characteristic caustics and the cusps in the lensing phenomena is largely expected. Besides the structure formation, the topological defects may be relevant to the baryon  asymmetry of the universe~\cite{Davis:1996zg}. The superconducting cosmic strings \cite{Witten:1984eb} or the stable loops of current-carrying string called the vortons~\cite{Carter:1999an}, may be responsible for the production of the high energy cosmic rays \cite{Bonazzola:1997tk} or even  the gamma ray bursts \cite{Berezinsky:2001cp}.

The vacuum polarisation effects in the presence of a cosmic string  is a much cultivated topic, both in flat and (anti)-de Sitter backgrounds. 
The study of the vacuum expectation values of the field squared, $\langle{0}|\phi^2(x)|{0}\rangle$ of a scalar field and of the fermionic condensate, $\langle{0}|\overline{\psi}(x){\psi}(x)|{0}\rangle$, the vacuum expectation values of their energy-momentum tensors as well as of  the conserved currents and various anomalies have  extensively been investigated in~\cite{Bellucci:2020qsr}-\cite{Spinelly:2003zd}. 
Apart from the usual Minkowski, de Sitter/anti-de Sitter backgrounds, these studies also include spacetimes with compactified dimensions, and even the presence of a magnetic flux. Several of such studies have been made in higher dimensions as well.

In this paper we wish to compute the vacuum polarisation effects of the Dirac fermions induced by a pointlike global monopole located in the cosmological de Sitter spacetime. 
For example, the renormalised  Euclidean Green function and the vacuum expectation value of the energy-momentum tensor for a fermionic field in a  global monopole located in flat spacetime was computed in~\cite{BezerradeMello:2007dxm} in the context of the  braneworld scenario. The vacuum expectation value of the energy-momentum tensor for massless as well as  massive fermions  were computed in~\cite{BezerradeMello:1999ge, Saharian:2005br}. Similar investigations were carried out by considering nontrivial core structure instead of a pointlike global monopole in~\cite{BezerradeMello:2007vz}.  The ground state energy of a massive scalar field in this background  using the $\zeta$-function regularisation can be seen in~\cite{BezerradeMello:1999mk}. Finite temperature effects on the vacuum expectation values of the energy-momentum tensor of the massless spin-1/2 fermions can be seen in~\cite{Carvalho:2001gi, BezerradeMello:2002rj}.   We further refer our reader to various quantum field theoretic computations in such backgrounds including the higher dimensions and the Kaluza-Klein theory in~\cite{BezerradeMello:2001pg, Saharian:2003sp, CavalcantideOliveira:2003sb, BezerradeMello:2002rp, Spinelly:2002mt, BezerradeMello:2012nq, Barbosa:2008fk, BezerradeMello:2006kka, BezerradeMello:2006tjm,  Spinelly:2005pd, CavalcantideOliveira:2003ie}. While all the above computations are done  in flat backgrounds, the computation of renormalised vacuum expectation value of the two point function for a scalar field  as well the energy-momentum tensor in a de Sitter global monopole background can be seen in~\cite{BezerradeMello:2009ju}.

 The rest of the paper is organised as follows. In the next Section and \ref{A}, we present the derivation of the  Dirac modes in the de Sitter global monopole background in a closed form. Using these mode functions, we compute the fermionic vacuum condensate, $\langle 0| \overline{\Psi}\Psi | 0\rangle$, and the vacuum expectation value of the energy-momentum tensor for the fermionic field respectively in \ref{s3}, \ref{s4} and \ref{s5}. \ref{B} and \ref{C} contain some detail for these Sections. Finally we conclude in \ref{s6}. We shall work in $3+1$-dimensions with mostly negative signature of the metric, and will set $c=1=\hbar$ throughout. We shall chiefly follow the method involving the Abel-Plana summation formula, as described in e.g.~\cite{Braganca:2019ord, BezerradeMello:2010ci} in the context of the de Sitter cosmic string background. 

\section{The Dirac modes}\label{s2}
We shall present below the derivation of an orthonormal, complete set of  Dirac modes  in the de Sitter background with  a  pointlike global monopole defect.  The metric in the cosmological time reads,
\begin{eqnarray}
\label{metric1}
ds^2=dt^2-e^{2t/\alpha}\left[dr^2+\beta^2 r^2 d\Omega^2\right],
\end{eqnarray}
where $\alpha^{-2}=\Lambda/3$, $\Lambda$ being the cosmological constant.  The parameter $\beta$ is conventionally expressed as $\beta^2=(1-8\pi G\xi^2)$~\cite{Barriola:1989hx}, where $\xi$ stands for the symmetry breaking scale, which lies well below the Planck energy, $G^{-1/2}$. Thus $0< \beta^2\leq 1$, and we have a deficit in the solid angle,  $4\pi(1-\beta^2)$ in \ref{metric1}.

In terms of the conformal time, $\eta=-\alpha e^{-t/\alpha}$ ($-\infty <\eta < 0^-$), the above metric reads,
\begin{eqnarray}
\label{metric2}
ds^2=\frac{\alpha^{2}}{\eta^2} \left(d\eta^2-dr^2-\beta^2 r^2d\Omega^2\right)
\end{eqnarray}
The pontlike global monopole breaks the maximal symmetry of the de Sitter spacetime and introduces a curvature singularity at $r=0$~\cite{BezerradeMello:2009ju},
 $$R=\frac{12}{\alpha^2}+\frac{2\left(1-\beta^2\right)\eta^2}{\alpha^2\beta^2 r^2}$$
 Setting  $\eta/\alpha \to 1$ in \ref{metric1}  recovers the metric of the flat spacetime with a global monopole, which may interestingly also describe an effective metric produced in the superfluid $^3{\rm H}$e-A, but with a negative deficit in the solid angle~\cite{BezerradeMello:2007vz}. 
 \\

The four orthonormal Dirac modes in the above background are given by (see \ref{A} for detail),
\begin{eqnarray}
\label{psi7p}
\Psi^{(+)}_{\sigma j l m}(\eta,r,\theta,\phi)&=&\frac{\lambda {\sqrt \pi} e^{\pi m\alpha/2}}{2\beta \alpha^{3/2} \sqrt{r}}\left( {\begin{array}{cc}
  \eta^2 H^{(1)}_{1/2-im\alpha}(\lambda|\eta|){J_{\nu_\sigma}}(\lambda r)\Omega_{jl_\sigma m}\\
   -i(\hat{r}\cdot{\vec{\sigma}}) \eta^2 H^{(1)}_{-1/2-im\alpha}(\lambda|\eta|){J_{\nu_\sigma+(-1)^\sigma}}(\lambda r)\Omega_{jl_\sigma m} \\
  \end{array} } \right) \nonumber\\
  \Psi^{(-)}_{\sigma j l m}(\eta,r,\theta,\phi)&=&\frac{\lambda {\sqrt \pi} e^{-\pi m\alpha/2}}{2\beta \alpha^{3/2}\sqrt{r}}\left( {\begin{array}{cc}
  i(-1)^\sigma(\hat{r}\cdot{\vec{\sigma}})\eta^2 H^{(2)}_{-1/2+im\alpha}(\lambda|\eta|){J_{\nu_\sigma+(-1)^\sigma}}(\lambda r)\Omega_{jl_\sigma m}\\
   \eta^2H^{(2)}_{1/2+im\alpha}(\lambda|\eta|){J_{\nu_\sigma}}(\lambda r)\Omega_{jl_\sigma m} \\
  \end{array} } \right)
\end{eqnarray}
where $\sigma =0,1$ correspond to two different solutions and the $(\pm)$-sign  in the superscripts represent respectively positive and negative frequency solutions. The ${\vec \sigma} $'s are the Pauli matrices. The $\Omega$'s are the spin-1/2 spherical harmonics and $l_\sigma$, $\nu_\sigma$ are respectively given in \ref{sigma7} and below \ref{rad_sol1}. $\lambda$ is a real, positive parameter.

It is easy to check that the above modes  satisfy 
\begin{eqnarray}
\left(\Psi^{(+)}_{\sigma j l m},\Psi^{(+)}_{\sigma' j' l' m'}\right)=\left(\Psi^{(-)}_{\sigma j l m},\Psi^{(-)}_{\sigma' j' l' m'}\right)=\delta_{jj'}\delta_{ll'}\delta_{mm'}{\delta_{\sigma\sigma'}}
\end{eqnarray}
with rest of the inner products vanishing. Thus they form a complete and orthonormal set.

\section{The fermionic condensate, $\langle 0| \overline{\Psi} \Psi | 0\rangle$}\label{s3}
Using the complete, orthonormal mode functions of \ref{psi7p}, we can quantise the fermionic field in the de Sitter-global monopole background.  Using this field quatisation, we wish to compute below the fermionic condensate, i.e. the vacuum expectation value of the operator  $ \overline{\Psi}\Psi$. By the mode-sum formula (e.g.~\cite{BezerradeMello:2009ng} and references therein), we have
\begin{eqnarray}
\label{conds}
\langle{0}|\overline{\Psi}\Psi |0 \rangle=\sum_{\sigma=0,1}\sum_{l,j,m}\int_{0}^{\infty} d\lambda\, {\overline{\Psi}^{(-)}_{\sigma jlm}\left(x\right)\Psi^{(-)}_{\sigma jlm}(x)}
\end{eqnarray}
where $\overline{\Psi}^{(\pm)}$=$\Psi^{\dagger(\pm)}\gamma^{(0)}$ is the  adjoint spninor. Using the second of \ref{psi7p} and the completeness relationship for the spherical harmonics, \ref{conds} can be expanded as
\begin{eqnarray}
\label{conds1}
\langle{0}|\overline{\Psi}\Psi |0 \rangle&=&\frac{\eta^4 e^{-m\pi\alpha}}{16\beta^2 \alpha^{3}r}\sum_{j=1/2}^{\infty}(2j+1)\int_{0}^{\infty} d\lambda \lambda^2\left(J_{\nu_1}^2(\lambda r)+J_{\nu_0}^2(\lambda r)\right)\left(|H^{(2)}_{1/2-im\alpha}\left(\lambda|\eta|\right)|^2-|H^{(2)}_{-1/2-im\alpha}\left(\lambda|\eta|\right)|^2\right)\nonumber\\
\end{eqnarray}
where $\nu_{\sigma}\,\,(\sigma =0,1)$ is given below \ref{rad_sol1}. In order to evaluate  the above integral, we use the following identities e.g.~\cite{BezerradeMello:2010ci},
\begin{eqnarray}
\label{relations}
|H^{(2)}_{\pm1/2-im\alpha}(\lambda|\eta|)|^2&=&\frac{4}{\pi^2}e^{m\pi\alpha}|K_{1/2\mp im\alpha}(i\lambda|\eta|)|^2,\;\;K_{1/2-im\alpha}(x)=K_{-1/2+im\alpha}(x)\nonumber\\
|K_{1/2-im\alpha}(i\lambda|\eta|)|^2-|K_{1/2+im\alpha}(i\lambda|\eta|)|^2&=&-\frac{i}{\lambda}\left(\partial_{|\eta|}+\frac{1-2im\alpha}{|\eta|}\right)K_{1/2-im\alpha}(i\lambda|\eta|)K_{1/2-im\alpha}(-i\lambda|\eta|) \qquad 
\end{eqnarray}
where $K$ is the modified Bessel function of the second kind. Then \ref{conds1} takes the form
\begin{eqnarray}
\label{conds2}
\langle{0}|\overline{\Psi}\Psi |0 \rangle =\frac{-i\eta^3}{4\pi^2\beta^2\alpha^{3} r}\sum_{j=1/2}^{\infty}(2j+1)\int d\lambda \lambda^2\left(J_{\nu_1}^2(\lambda r)+J_{\nu_0}^2(\lambda r)\right)
\left(\eta\partial_{\eta}+{1-2im\alpha}\right)K_{1/2-im\alpha}(i\lambda|\eta|)K_{1/2-im\alpha}(-i\lambda|\eta|)
\end{eqnarray}
In order to evaluate the above integral we need to simplify the product $K_{1/2-im\alpha}(i\lambda|\eta|)K_{1/2-im\alpha}(-i\lambda|\eta|)$. We write it as an integral representation \cite{Prud, Grad, Wats44, Abra64}
\begin{eqnarray}
\label{integral_1}
K_{1/2-im\alpha}(i\lambda|\eta|)K_{1/2-im\alpha}(-i\lambda|\eta|)
&=& \int_0^\infty \frac{du}{u}\int_{0}^\infty dy'\;\cosh 2\mu y' \,e^{-2u \lambda^2\eta^2\sinh^2y'-1/2u},\;\mu=1/2-im\alpha \qquad
\end{eqnarray} 
Substituting \ref{integral_1} into \ref{conds2},  we have
\begin{eqnarray}
\label{conds31}
\langle{0}|\overline{\Psi}\Psi |0 \rangle&=&\frac{-i\eta^3}{4\pi^2\beta^2\alpha^{3} r}\left(\eta\partial_{\eta}+{1-2im\alpha}\right)\sum_{j=1/2}^{\infty}(2j+1)\int_0^\infty \frac{du}{u}e^{-1/2u}\int_{0}^\infty dy'\;\cosh2\mu y'\nonumber\\
&&\times \int d\lambda \lambda\left(J_{\nu_1}^2(\lambda r)+J_{\nu_0}^2(\lambda r)\right)e^{-2\lambda^2u\eta^2\sinh^2y'}\nonumber\\
&=&\frac{-i\eta^3}{4\pi^2\beta^2\alpha^{3} r}\left(\eta\partial_{\eta}+{1-2im\alpha}\right)\sum_{j=1/2}^{\infty}(2j+1)\int_{0}^\infty dy'\,\cosh2\mu y'\;\int_0^\infty \frac{du}{u}\frac{e^{-1/2u}}{4u\eta^2\sinh^2{y'}}e^{-r^2/(4u\eta^2\sinh^2{y'})}\nonumber\\
&&\times \left[I_{\nu_1}(r^2/4u\eta^2\sinh^2{y'})+I_{\nu_0}(r^2/4u\eta^2\sinh^2{y'})\right],
\end{eqnarray}
where $I_{\nu}$ is the modified Bessel function of the first kind. We introduce a new variable $x=r^2/(4u\eta^2\sinh^2{y'})$ and perform the $y'$-integral in \ref{conds31} to have
\begin{eqnarray}
\label{conds3}
\langle{0}|\overline{\Psi}\Psi |0 \rangle&=&\frac{-i\eta^3}{8\pi^2\beta^2\alpha^{3} r^3}\left(\eta\partial_\eta+{1-2im\alpha}\right)\sum_{j=1/2}^{\infty}(2j+1)\int_0^\infty {dx}\;e^{-x(1-\eta^2/r^2)}{\left(I_{\nu_1}(x)+I_{\nu_0}(x)\right)}K_{1/2-im\alpha}(x\eta^2/r^2) \nonumber\\
\end{eqnarray}
We next use~\cite{BezerradeMello:2010ci}
$$\left(\eta\partial_{\eta}+{1-2im\alpha}\right)e^{x\eta^2/r^2}K_{1/2-im\alpha}(x\eta^2/r^2)=\frac{2 x\eta^2}{r^2}e^{x\eta^2/r^2}\left(K_{1/2-im\alpha}(x\eta^2/r^2)-K_{-1/2-im\alpha}(x\eta^2/r^2)\right),$$
in order  to further simplify \ref{conds3} as
\begin{eqnarray}
\label{conds4'}
\langle{0}|\overline{\Psi} \Psi |0 \rangle&=&\frac{\eta }{2\pi^2\beta^2 \alpha^{3}r}\sum_{j=1/2}^{\infty}(2j+1)\int_0^\infty {dy}\;y\;e^{y\left(1-r^2/\eta^2\right)}\text{Im}[K_{1/2-im\alpha}(y)]\left(I_{\nu_1}(yr^2/\eta^2)+I_{\nu_0}(yr^2/\eta^2)\right)
\qquad
\end{eqnarray}
where following~\cite{BezerradeMello:2010ci}, we also have defined a new variable, $y= x \eta^2/r^2$. We wish to split now $\langle{0}|\overline{\Psi}\Psi|{0}\rangle$ into two parts -- one corresponding to the pure de Sitter spacetime without any monopole ($\beta=1$) and the other corresponding to the global monopole defect,  so that we {\it define}
\begin{eqnarray}
\label{conds4}
\langle{0}|\overline{\Psi}\Psi|{0}\rangle_{\text{gm}}:= \langle{0}|\overline{\Psi} \Psi|{0}\rangle-\langle{0}|\overline{\Psi}\Psi|{0}\rangle_{\text{dS}}
\end{eqnarray}
 Similar things can be seen in e.g.~\cite{BezerradeMello:2010ci} and references therein in the context of cosmic strings. For the pure de Sitter part, we have the renormalised expression,
\begin{eqnarray}
\label{final_2}
\langle{0}|\overline{\Psi}\Psi|{0}\rangle_{\text{dS,~Ren.}}=\frac{m }{2\pi^2\alpha^{2}}\left(1+m^2\alpha^2\right)\left[\ln\left(m\alpha\right)-\text{Re}\, \psi(im\alpha)+\frac{1}{12}\right],
\end{eqnarray}
where $\psi$ is the digamma function~\cite{Abra64}. Although the above expression has been found earlier in  various references (e.g.~\cite{BezerradeMello:2010ci} in the context of de Sitter cosmic strings),  we have outlined its derivation in \ref{B}, as a check of consistency of our modes, \ref{psi7p}. A couple of comments  regarding the definition of \ref{conds4} are in order here. It will turn out below that the quantity $\langle{0}|\overline{\Psi}\Psi|{0}\rangle_{\text{gm}}$ is ultraviolet finite. In other words, the curvature induced by the defect $\beta<1$ in  \ref{metric1} does not induce any divergence. The divergence only comes from the pure de Sitter part and can be renormalised via a cosmological constant counterterm.   This  seems to  have some qualitative similarity (but not the same) with the Hadamard subtraction  cum regularisation  of the Feynman propagator in a general curved spacetime using the Riemann normal coordinates, e.g.~\cite{Parker:2009uva}.   One can try regularisation techniques different from that of the present one, as well. Apart from the left hand side being regular, the decomposition of  \ref{conds4} has the obvious advantage regarding the application of the Abel-Plana  summation formula, eventually helping us to do many  computations analytically, we will see below. Despite this advantage, however, this regularisation scheme might have some caveat as well, as we shall point out in \ref{s6}.    \\

 After subtracting the integral expression of $\langle{0}|\overline{\Psi}\Psi|{0}\rangle_{\text{dS}}$, \ref{dsgms-1},  from \ref{conds4'},  it turns out from \ref{conds4} that we need to evaluate the expression,
 $$\sum_{j}(j+1/2)\left[\left(\frac{1}{\beta^2}I_{\frac{\left(j+1/2\right)}{\beta}+ 1/2}-I_{j+1/2+1/2}\right)+\left(\frac{1}{\beta^2}I_{\frac{\left(j+1/2\right)}{\beta}- 1/2}-I_{j+1/2-1/2}\right)\right],$$
which can be done conveniently by using the Abel-Plana summation 
formula~\cite{Saharian:2007ph},
\begin{eqnarray}
\label{abel-planna1}
\sum_{j=0}^\infty 
f(j)=\frac{f(0)}{2}+\int_{0}^{\infty}dv f(v)-i\int_{0}^{\infty}du\frac{\left(f(iu)-f(-iu)\right)}{e^{2\pi u}+1}
\end{eqnarray}
Thus it follows from the above equation that
\begin{eqnarray}
\label{abel-planna2}
\sum_{j}\left[f(j/\beta)/\beta-f(j)\right]=-i\int_{0}^{\infty}du{\left(f(iu)-f(-iu)\right)}\left(\frac{1}{e^{2\pi u/\beta}+1}-\frac{1}{e^{2\pi u}+1}\right)
\end{eqnarray}
When we apply this formula in our case we have the relevant integral to be 
\begin{eqnarray}
\label{abel-planna3}
&&\sum_{j=1/2}^{\infty}(j+1/2)\Bigg[\frac{1}{\beta^2}I_{\frac{j+1/2}{\beta}+1/2}(yr^2/\eta^2)-I_{{j+1/2}+1/2}(yr^2/\eta^2)+\frac{1}{\beta^2}I_{\frac{j+1/2}{\beta}-1/2}(yr^2/\eta^2)-I_{{j+1/2}-1/2}(yr^2/\eta^2)\Bigg]\nonumber\\
&=&-\frac{4}{\pi}\int du\, u \,g(\beta,u)\text{Im}\left[K_{1/2+iu}(yr^2/\eta^2)\right]
\end{eqnarray}
where we have defined,
\be
g(\beta,u)=\cosh\pi u\left(\frac{1}{e^{2\pi\beta u}+1}-\frac{1}{e^{2\pi u}+1}\right)
\label{g}
\ee
Using this,  we have after some algebra
\begin{eqnarray}
\label{gmconds1}
\langle{0}|\overline{\Psi}\Psi|{0}\rangle_{\text{gm}}&&= \langle{0}|\overline{\Psi}\Psi|{0}\rangle-\langle{0}|\overline{\Psi}\Psi|{0}\rangle_{\text{dS}}\nonumber\\ &&= -\frac{4\eta }{\pi^3 \alpha^{3} r}\int_0^\infty\;du\;u\;g(\beta,u)\int_0^\infty {dy}\;y\;e^{y\left(1-r^2/\eta^2\right)}\text{Im}[K_{1/2-im\alpha}(y)]\text{Im}[K_{1/2+iu}(yr^2/\eta^2)]
\qquad
\end{eqnarray}

We shall evaluate the above integral numerically. However,  two special cases are of interest, for which analytic expressions can be found. The first is when the proper distance from the monopole $(r=0)$ is small, $r\alpha/|\eta| \to 0$. Recalling $y= x \eta^2 /r^2$ (cf., the discussion below \ref{conds4'}),  we have for large $y$-values~\cite{Abra64} 
 $$\text{Im}[K_{1/2-im\alpha}(y)] \approx -\frac{m\sqrt{\pi}\alpha}{(2y)^{3/2}}e^{-y},$$
 so that \ref{gmconds1} becomes
\begin{eqnarray}
\label{gmconds2}
\langle{0}|\overline{\Psi}\Psi|{0}\rangle_{\text{gm}}&=&\frac{2^{1/2}\eta m}{\pi^{5/2} \alpha^{2}r}\int_0^\infty\;du\;u\;g(\beta,u)\int_0^\infty {dy}\;y^{-1/2}\;e^{-yr^2/\eta^2}\text{Im}[K_{1/2+iu}(yr^2/\eta^2)]
\end{eqnarray}
Using now the integral~\cite{Grad} 
\begin{eqnarray}
\label{identity1}
\int_0^\infty {d\zeta } \zeta^{\beta-1} e^{-\zeta} K_{\nu}(\zeta)=\frac{\sqrt{\pi}\Gamma(\beta+\nu)\Gamma(\beta-\nu)}{2^\beta\Gamma(\beta+1/2)}
\end{eqnarray}
\ref{gmconds2} becomes
\begin{eqnarray}
\label{gmconds4}
{\langle{0}|\overline{\Psi}\Psi|{0}\rangle_{\text{gm}}}\Big\vert_{r/|\eta| \to 0}&=&\frac{m \eta^2}{\pi^{2}  \alpha^2 r^2}\text{Im}
\int_0^\infty\;du\;u\;g(\beta,u){\Gamma(1+iu)\Gamma(-iu)}\nonumber\\
&=&\frac{m }{24\pi\left(r \alpha/\eta\right)^2}\left(\frac{1}{\beta^2}-1\right)+\frac{m}{2\pi^3\left(r \alpha/\eta\right)^2}\sum_{n=0}^\infty (-1)^n \left({\zeta(2,1+(n+1)\beta)}-{\zeta(2,1+(n+1))}\right)\nonumber\\
\end{eqnarray}
where we have used $\Gamma(1-iu)\Gamma(iu)=i\pi/\sinh{\pi u}$~\cite{Abra64} and the expansion,
 \begin{eqnarray} 
 \label{express1}
\frac{g(\beta,u)}{\sinh{\pi u}}
=\left(\frac{1}{e^{2\pi\beta u}+1}-\frac{1}{e^{2\pi u}+1}\right)+\frac{2}{e^{2\pi u}-1}\sum_{n=0}(-1)^{n}\left(e^{-2\pi(n+1)\beta u}-e^{-2\pi(n+1)u}\right)
\end{eqnarray}
In order to obtain \ref{gmconds4}, we also have used the  formula
given in \cite{Grad}  involving the multiple $\zeta$-function  
$$\zeta(2,1+(n+1)\beta)=\sum_{n_1=1}^\infty\frac{1}{(n_1+1)^2}\sum_{n_2=1}^{n_1}\frac{1}{(n_1+1)^{1+(n+1)\beta}}$$
The above series is obviously convergent, as $\beta>0$. For example, for $\beta=0.5$, its numerical value is close to $0.9$, as can be easily checked using e.g., Mathematica.  

Thus,  the fermionic condensate \ref{gmconds4} diverges  as the square of the proper radial distance $r\alpha/|\eta|$, as we move towards the monopole, expected due to the curvature singularity at $r=0$. Note also that the condensate vanishes for $m=0$.
 \\

 As the second special case, let us consider the opposite scenario, i.e. large proper distance from the monopole, $r\alpha/|\eta| \gg 1$. Since $y= x \eta^2/r^2$ (cf., the discussion below \ref{conds4'}), we make an expansion for small $y$-values~\cite{Grad}
$$K_{1/2-i m\alpha}(y)\approx  \frac{1}{2}\Gamma\left(\frac12-i m\alpha\right)\left(\frac{y}{2}\right)^{i m\alpha-1/2}$$
Substituting the above into \ref{gmconds1} and using \ref{identity1}, we find
\begin{eqnarray}
\label{gmconds5}
\langle{0}|\overline{\Psi}\Psi|{0}\rangle_{\text{gm}}\Big\vert_{r/|\eta| \to \infty}&\approx&-\frac{4\eta }{\pi^3 \alpha^{3}r}\int_0^\infty du\, u\ g(\beta,u)\int_0^\infty {dy} y e^{-yr^2/\eta^2}\text{Im}\Bigg[\frac{1}{2}\Gamma(1/2-i m\alpha)\left(\frac{y}{2}\right)^{i m\alpha-1/2}\text{Im}[K_{1/2+iu}(yr^2/\eta^2)]\Bigg]\nonumber\\
&=&\frac{1}{2^{1/2}\pi^{7/2}\alpha^3\left(r/\eta\right)^{4}}\int_0^\infty\;du\;u^2\;g(\beta,u)\Bigg[\frac{\Gamma(1/2-i m\alpha)}{\Gamma(2+i m\alpha)}\left(\frac{\eta}{2r}\right)^{2 mi\alpha}\Gamma(1+i m\alpha+i u)\Gamma(1+i m\alpha-i u)\Bigg]\nonumber\\
&=&\frac{2^{1/2}\alpha f(\beta,m\alpha)}{\pi^{7/2}\left(r\alpha/\eta\right)^{4}}\sin\left(2m\alpha\,\text{ln}\left(2r/|\eta|\right)-\varphi_0\right)
\end{eqnarray}
showing quartic fall off along with oscillatory behaviour. The phase $\varphi_0$ and the function $f(\beta,m\alpha)$  above  are determined by the complex relationship
\begin{eqnarray}
\label{B-function}
f(\beta,m\alpha)e^{i\varphi_0}=\frac{\Gamma(1/2-i m\alpha)}{\Gamma(2+i m\alpha)}\int_0^\infty du u^2 g(\beta,u) \Gamma(1+i m\alpha+i u)\Gamma(1+i m\alpha-i u)
\end{eqnarray}
Note that if we set $m=0$ above, the integral becomes real and hence $\phi_0$ becomes vanishing. This makes \ref{gmconds5} vanishing too, in the massless limit. We also note that the radial dependence of the divergence in \ref{gmconds4}  or the fall off in \ref{gmconds5}, are qualitatively similar to that of the de Sitter cosmic string ~\cite{BezerradeMello:2010ci}.

\begin{figure}[h]
\includegraphics[scale=0.70]{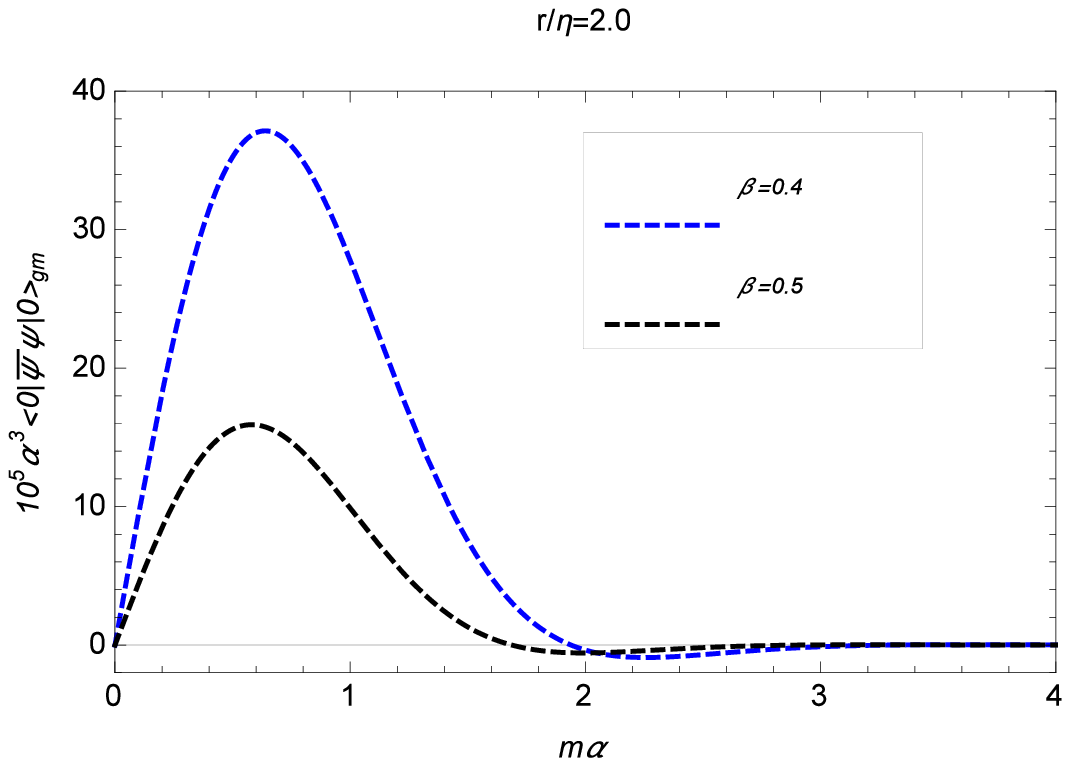}\hspace{0.2cm}
\includegraphics[scale=0.7]{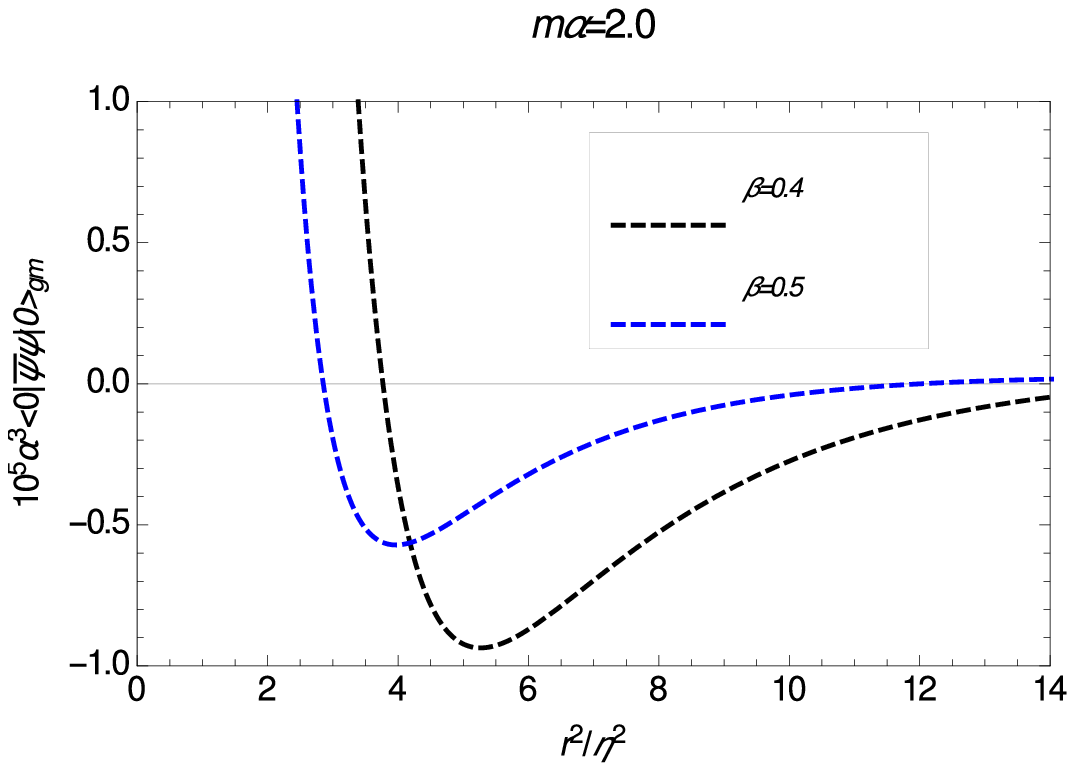}
\caption{Global monopole induced part of the vacuum expectation value, $\langle{0}|\overline{\Psi}\Psi|{0}\rangle_{\text{gm}}$, \ref{gmconds1}, as a function of the dimensionless mass parameter $m\alpha$ (left) and as a function of the dimensionless proper radial distance squared, $r^2/\eta^2$. See main text for discussion.}
\label{f1}
\end{figure}

Finally, we have plotted the variation of the condensate $\langle{0}|\overline{\Psi}\Psi|{0}\rangle_{\text{gm}}$, \ref{gmconds1}, with respect to the dimensionless mass parameter $m\alpha$ and the dimensionless proper distance squared $r^2/\eta^2$, in \ref{f1}, for two different values of the defect parameter $\beta$.  For sufficiently high values of either of these variables, the curves tend to merge. The first plot shows that the condensate is vanishing for $m=0$, in agreement with our previous results found for the special cases.  
%

\section{Vacuum expectation value of the energy-momentum tensor}\label{s4}

We next wish to compute the vacuum expectation value of the fermionic energy-momentum tensor in the background \ref{metric2}. We have for the Dirac field,
\begin{eqnarray}
\label{emt1}
T_{\mu\nu}=\frac{i}{2}\left[{\overline{\Psi}}\gamma_{(\mu}\nabla_{\nu)}{\Psi}-(\nabla_{(\mu}{\overline{\Psi}})\gamma_{\nu)}{\Psi}\right]
\end{eqnarray}
Expanding the spinor  and its adjoint in terms of mode functions, and taking the expectation value with respect to the vacuum, we have
\begin{eqnarray}
\langle 0|T_{\mu\nu}|0\rangle=\frac{i}{2}\int_{0}^{\infty} d\lambda\sum_{\sigma,j,l,m}\left[{\overline{\Psi}^{(-)}_{\sigma j l m}}(x)\gamma_{(\mu}\nabla_{\nu)}{\Psi^{(-)}_{\sigma j l m}(x)-(\nabla_{(\mu}\overline{\Psi}^{(-)}_{\sigma j l m}})\gamma_{\nu)}{\Psi^{(-)}_{\sigma j l m}}(x)\right]
\label{T1}
\end{eqnarray}
In the above expression we  encounter anti-commutators between the $\gamma$'s and the spin connection marices, $[\gamma_{\mu}, \Gamma_{\nu}]_+$. However, using \ref{basis} and \ref{gamma2'}, it is easy to see that such anti-commutators vanish for all $\mu$, $\nu$, leaving us only with the partial derivatives to deal with,
\begin{eqnarray}
\langle 0|T_{\mu\nu}|0\rangle=\frac{i}{2}  \int_{0}^{\infty} d\lambda\sum_{\sigma,j,l,m}\left[{\overline{\Psi}^{(-)}_{\sigma j l m}}(x)\gamma_{(\mu}\partial_{\nu)}{\Psi^{(-)}_{\sigma j l m}(x)-(\partial_{(\mu}\overline{\Psi}^{(-)}_{\sigma j l m}})\gamma_{\nu)}{\Psi^{(-)}_{\sigma j l m}}(x)\right]
\label{T2}
\end{eqnarray}
Using now the second of \ref{psi7p}, we shall explicitly compute below the above expression component-wise.
\subsection{Energy density}

Using \ref{basis} in order to express the curved space $\gamma$-matrices in terms of the flat space ones, we have from \ref{T2}
\begin{eqnarray}
\label{emt_2}
\langle{0}|T^0_{0}|0 \rangle&=&\frac{i\eta}{2 \alpha}\int_{0}^{\infty} d\lambda\sum_{\sigma j l m}\left[{{\Psi^\dagger}^{(-)}_{\sigma j l m}}(x)\partial_{0}{\Psi^{(-)}_{\sigma j l m}}(x)-(\partial_{0}{\Psi^\dagger}^{(-)}_{\sigma j l m}(x)){\Psi^{(-)}_{\sigma j l m} }(x)\right]
\end{eqnarray}
Using now the second of \ref{psi7p} and the formula for the spherical harmonics, 
\begin{eqnarray}
\label{spherical_harmonics}
\sum_{m_j=-j}^{m_j=+j}|\Omega_{jlm}|^2=\frac{2j+1}{4\pi},
\end{eqnarray}
\ref{emt_2} takes the form after some algebra,
\begin{eqnarray}
\label{emt_3}
\langle 0|T^{0}_{0}|0\rangle&=&\frac{\eta^5 }{4\pi^2\beta^2 \alpha^{4} r}\sum_{j=1/2}^{\infty}(2j+1)\int d\lambda \lambda \left(J_{\frac{j+1/2}{\beta}+1/2}^2(\lambda r)+J_{\frac{j+1/2}{\beta}-1/2}^2(\lambda r)\right)\nonumber\\
&&\Bigg[\Bigg(\partial_{\eta}^2+\frac{2}{\eta}\partial_{\eta}+4\left(\lambda^2+\frac{i m\alpha(1/2-im\alpha)}{\eta^2}\right)\Bigg)K_{1/2-im\alpha}(i\lambda|\eta|)K_{1/2-im\alpha}(-i\lambda|\eta|)\Bigg]
\end{eqnarray}
Rearranging the above expression a little bit, we have
\begin{eqnarray}
\label{vev_00comp}
\langle{0}|T^0_0|{0}\rangle &=&\frac{\eta^5 }{4\pi^2\beta^2 \alpha^{4} r}\sum_{j=1/2}^{\infty}(2j+1)\Bigg[\left(\partial_{\eta}^2+\frac{2}{\eta}\partial_{\eta}+\frac{4i m\alpha(1/2-im\alpha)}{\eta^2}\right)\nonumber\\
&&\times\int d\lambda \lambda \left(J_{\frac{j+1/2}{\beta}+1/2}^2(\lambda r)+J_{\frac{j+1/2}{\beta}-1/2}^2(\lambda r)\right)K_{1/2-im\alpha}(i\lambda|\eta|)K_{1/2-im\alpha}(-i\lambda|\eta|)\nonumber\\
&&+4\int d\lambda \lambda^3 \left(J_{\frac{j+1/2}{\beta}+1/2}^2(\lambda r)+J_{\frac{j+1/2}{\beta}-1/2}^2(\lambda r)\right)K_{1/2-im\alpha}(i\lambda|\eta|)K_{1/2-im\alpha}(-i\lambda|\eta|)\Bigg]
\end{eqnarray}

Using the integral representation for the product $K_{1/2-im\alpha}(i\lambda|\eta|)K_{1/2-im\alpha}(-i\lambda|\eta|)$, \ref{integral_1}, we reduce the first integral in \ref{vev_00comp} to
\begin{eqnarray}
\label{first_int}
&&\int d\lambda \lambda \left(J_{\frac{j+1/2}{\beta}+1/2}^2(\lambda r)+J_{\frac{j+1/2}{\beta}-1/2}^2(\lambda r)\right)K_{1/2-im\alpha}(i\lambda|\eta|)K_{1/2-im\alpha}(-i\lambda|\eta|)\nonumber\\
&=&\frac{1}{2r^2}\int_0^\infty{dx}e^{-x}\left(I_{\frac{j+1/2}{\beta}+1/2}(x)+I_{\frac{j+1/2}{\beta}-1/2}(x)\right)e^{x\eta^2/r^2}K_{1/2-im\alpha}(x\eta^2/r^2),
\end{eqnarray}
where for the variable $x$ we used the previous definition $x=r^2/(4u\eta^2\sinh^2{y^\prime})$ (cf., the discussion below \ref{conds31}) and have performed the integration over $y'$. \\

Similarly, for the second integral in \ref{vev_00comp}, we have
\begin{eqnarray}
\label{second_int}
&&\int d\lambda \lambda^3 \left(J_{\frac{j+1/2}{\beta}+1/2}^2(\lambda r)+J_{\frac{j+1/2}{\beta}-1/2}^2(\lambda r)\right)K_{1/2-im\alpha}(i\lambda|\eta|)K_{1/2-im\alpha}(-i\lambda|\eta|)\nonumber\\
&=&-\frac{2}{r^4}\frac{\partial}{\partial{\eta^2}}\eta^2\int_0^\infty\;dx\;x\;{e^{-x}}\left(I_{\frac{j+1/2}{\beta}+1/2}(x)+I_{\frac{j+1/2}{\beta}-1/2}(x)\right)e^{x\eta^2/r^2}K_{1/2-im\alpha}(x\eta^2/r^2)
\end{eqnarray}
As earlier, by defining $x=yr^2/\eta^2$, we rewrite the differential operator appearing in \ref{vev_00comp} as, 
\begin{eqnarray}
\label{operator3}
\partial_{\eta}^2+\frac{2 \partial_{\eta}} {\eta}+\frac{4im\alpha\left(1/2-im\alpha\right)}{\eta^2}\equiv \frac{4x}{r^2}\left(x\partial_x^2+\frac{3 \partial_x}{2}+ \frac{im\alpha\left(1/2-im\alpha\right)}{x}\right)
\end{eqnarray}
Substituting now \ref{first_int}, \ref{second_int} and \ref{operator3} into \ref{vev_00comp}, we have the vacuum expectation value  of the energy density,
\begin{eqnarray}
\label{energy_density}
\langle{0}|T^0_0|{0}\rangle&=&\frac{\eta^5}{2\pi^3 \beta^2  \alpha^{4} {r^5}}\sum_{j=1/2}^{\infty}(2j+1)\int_0^\infty{dx}\;x\;e^{-x}\left(I_{\frac{j+1/2}{\beta}+1/2}(x)+I_{\frac{j+1/2}{\beta}-1/2}(x)\right)\nonumber\\
&&\times \left(x\partial_x^2+\left(\frac32-2x\right)\partial_x-2+\frac{im\alpha\left(1/2-im\alpha\right)}{x}\right)e^{x\eta^2/r^2}K_{1/2-im\alpha}(x\eta^2/r^2)
\end{eqnarray}
Following now~\cite{BezerradeMello:2010ci}, we use the properties of $K_{1/2-im\alpha}$ to simplify the operator part of the above equation as 
\begin{eqnarray}
&& \left(x\partial_x^2+\left(\frac32-2x\right)\partial_x-2+\frac{im\alpha\left(1/2-im\alpha\right)}{x}\right)e^{x\eta^2/r^2}K_{1/2-im\alpha}(x\eta^2/r^2)\nonumber\\
&=&-\frac{1}{2}e^{x\eta^2/r^2}\left(K_{1/2-im\alpha}(x\eta^2/r^2)+K_{-1/2+im\alpha}(x\eta^2/r^2)\right)
\end{eqnarray}
Using the above expression and also once again converting the variable to $y=x\eta^2/r^2$ for our convenience, the integral of \ref{energy_density} takes the form,
\begin{eqnarray}
\label{energy-density3}
\langle{0}|T^0_0|{0}\rangle&=&\frac{\eta }{4\pi^2\beta^2 \alpha^{4} {r}}\sum_{j=1/2}^{\infty}(2j+1)\int_0^\infty{dy}\;y\;e^{y\left(1-r^2/\eta^2\right)}\left(I_{\frac{j+1/2}{\beta}+1/2}(yr^2/\eta^2)+I_{\frac{j+1/2}{\beta}-1/2}(yr^2/\eta^2)\right)\text{Re}[K_{1/2-im\alpha}(y)],\nonumber\\
\end{eqnarray}
%

\subsection{Radial pressure}

We have
\begin{eqnarray}
\label{rr-comp}
\langle{0}|T_{rr}|0\rangle=\frac{i}{2}\int d\lambda \sum_{\sigma j l m}\left[{\Psi^\dagger}^{(-)}_{\sigma j l m}\gamma^{(0)}\gamma_{r}\partial_{r}\Psi^{(-)}_{\sigma j l m}-(\partial_{r}{\Psi^\dagger}^{(-)}_{\sigma j l m})\gamma^{(0)}\gamma_{r}\Psi^{(-)}_{\sigma j l m}\right]
\end{eqnarray}
which,  after using \ref{psi7p} and \ref{basis} becomes
\begin{eqnarray}
\label{radial_1}
\langle{0}|T^{r}_{r}|{0}\rangle&=&\frac{\eta^5 }{4\pi^2\beta^2 \alpha^{4} r}\sum_{j=1/2}^{\infty}(2j+1)\int d\lambda \lambda^3 \Bigg(J_{\frac{j+1/2}{\beta}-1/2}(\lambda r)\partial_r{J_{\frac{j+1/2}{\beta}+1/2}}(\lambda r)-J_{\frac{j+1/2}{\beta}+1/2}(\lambda r)\partial_r{J_{\frac{j+1/2}{\beta}-1/2}(\lambda r)}\Bigg)\nonumber\\ 
&&\times \left(K_{1/2-im\alpha}(i\lambda|\eta|)K_{1/2-im\alpha}(-i\lambda|\eta|)+{\rm c.c.}\right)
\end{eqnarray}

In order to evaluate the integral in \ref{radial_1}, we further use~\cite{BezerradeMello:2010ci}
\begin{eqnarray}
\label{formula_1}
&&J_{\frac{j+1/2}{\beta}-1/2}(\lambda r)\partial_r{J_{\frac{j+1/2}{\beta}+1/2}}(\lambda r)-J_{\frac{j+1/2}{\beta}+1/2}(\lambda r)\partial_r{J_{\frac{j+1/2}{\beta}-1/2}(\lambda r)}\nonumber\\
&=&\frac{1}{\lambda^2}\left(\frac{1}{2}\partial_r^2+\frac{1}{r}\partial_r+\frac{j+1/2}{\beta}\left(\frac{1/2-\frac{j+1/2}{\beta}}{r^2}\right)\right)J_{{\frac{j+1/2}{\beta}-1/2}}^2(\lambda r)+2J_{{\frac{j+1/2}{\beta}-1/2}}^2(\lambda r)
\end{eqnarray}
Substituting this into \ref{radial_1}, we have
\begin{eqnarray}
\label{radial_2}
\langle{0}|T^{r}_{r}|{0}\rangle&=&-\frac{\eta^5 }{4\pi^2\beta^2 \alpha^{4} r}\sum_{j=1/2}^{\infty}(2j+1)\int d\lambda \Bigg(\lambda\left(\frac{1}{2}\partial_r^2+\frac{1}{r}\partial_r+\frac{j+1/2}{\beta}\left(\frac{1/2-\frac{j+1/2}{\beta}}{r^2}\right)\right)J_{{\frac{j+1/2}{\beta}-1/2}}^2(\lambda r)\nonumber\\
&&+2\lambda^3 J_{{\frac{j+1/2}{\beta}-1/2}}^2(\lambda r)\Bigg)\Bigg(K_{1/2-im\alpha}(i\lambda|\eta|)K_{1/2-im\alpha}(-i\lambda|\eta|)+{\rm c.c.}\Bigg)
\end{eqnarray}
Now the first term in the above  integral  can be rewritten as, after using the integral representation \ref{integral_1}
\begin{eqnarray}
\label{int4}
&&\int d\lambda \lambda J_{{\frac{j+1/2}{\beta}-1/2}}^2(\lambda r)\Bigg(K_{1/2-im\alpha}(i\lambda|\eta|)K_{1/2-im\alpha}(-i\lambda|\eta|)+{\rm c.c.}\Bigg)\nonumber\\
 &=&\frac{1}{2r^2}\int_0^\infty{dx}e^{-x}I_{\frac{j+1/2}{\beta}-1/2}(x)e^{x\eta^2/r^2}\left( K_{1/2-im\alpha}(x\eta^2/r^2)+{\rm c.c.}\right)
\end{eqnarray}

Likewise, the second integral of \ref{radial_2} can be recast as
\begin{eqnarray}
\label{int5}
&&\int d\lambda \lambda^3 J^2_{{\frac{j+1/2}{\beta}-1/2}}(\lambda r)\Bigg(K_{1/2-im\alpha}(i\lambda|\eta|)K_{1/2-im\alpha}(-i\lambda|\eta|)+{\rm c.c.}\Bigg)\nonumber\\
&=&\frac{2}{r^4}\int_0^\infty{dx}\left(\left(x\partial_x+1\right) e^{-x}I_{\frac{j+1/2}{\beta}-1/2}(x)\right)e^{x\eta^2/r^2}\left( K_{1/2-im\alpha}(x\eta^2/r^2)+{\rm c.c.}\right)
\end{eqnarray}
where in both the above cases  we have defined the variable $x=yr^2/\eta^2$, as earlier.

Using now \ref{int4} \ref{int5} and once again \ref{integral_1} into \ref{radial_2}, and converting the variables to $y=x\eta^2/r^2$, we find after a little algebra
\begin{eqnarray}
\label{rr-emt1}
\langle{0}|T^r_r|{0}\rangle&=&\frac{\eta }{4\pi^2\beta^2 \alpha^{4} r}\sum_{j=1/2}^\infty\left(2j+1\right)\int_0^\infty{dy}\;y\;e^{y\left(1-r^2/\eta^2\right)}\;\text{Re}[K_{1/2-im\alpha}(y)]\nonumber\\
&&\times \left(I_{\frac{j+1/2}{\beta}+1/2}(yr^2/\eta^2)+I_{\frac{j+1/2}{\beta}-1/2}(yr^2/\eta^2)\right)=\langle{0}|T^0_0|{0}\rangle
\end{eqnarray}
where the last equality follows from comparison with \ref{energy-density3}.

\subsection{The angular stresses}

We have
\begin{eqnarray}
\label{theta_1}
\langle{0}|T^{\theta}_{\theta}|{0}\rangle&=&\frac{\eta^5 }{4\pi^2\beta^3 \alpha^{4} r^2}\sum_{j=1/2}^{\infty}(2j+1)^2\int d\lambda \lambda^2 J_{\frac{j+1/2}{\beta}-1/2}(\lambda r)J_{\frac{j+1/2}{\beta}+1/2}(\lambda r)\nonumber\\ 
&&\times\left(K_{1/2-im\alpha}(i\lambda|\eta|)K_{1/2-im\alpha}(-i\lambda|\eta|)+{\rm c.c.}\right)
\end{eqnarray}
For the product of the two Bessel functions appearing above, we use the relation~\cite{Braganca:2019ord}
$$ J_{\frac{j+1/2}{\beta}-1/2}(\lambda r)J_{\frac{j+1/2}{\beta}+1/2}(\lambda r)=\frac{1}{\lambda}\left(\frac{(j+1/2)/\beta-1/2}{r}-\frac{1}{2}\partial_r\right)J_{\frac{j+1/2}{\beta}-1/2}^2(\lambda r)$$
We then have after using \ref{integral_1},
\begin{eqnarray}
\label{11-theta-theta-1}
&&\sum_{j=1/2}^{\infty}(2j+1)^2\int d\lambda \lambda \Bigg(\left(\frac{(j+1/2)/\beta-1/2}{r}-\frac{1}{2}\partial_r\right)J_{\frac{j+1/2}{\beta}-1/2}^2(\lambda r)\Bigg)\nonumber\\
&&\times \Bigg(K_{1/2-im\alpha}(i\lambda|\eta|)K_{1/2-im\alpha}(-i\lambda|\eta|)+{\rm c.c.}\Bigg)\nonumber\\
&=&\frac{r}{\eta^4}\sum_{j=1/2}^{\infty}(2j+1)^2\int_0^\infty{dy}\;y\;e^{y}\text{Re}\left[K_{1/2-im\alpha}(y)\right]e^{-yr^2/\eta^2}
\left(I_{\frac{j+1/2}{\beta}-1/2}(yr^2/\eta^2)-I_{\frac{j+1/2}{\beta}+1/2}(yr^2/\eta^2)\right)\nonumber\\
\end{eqnarray}
Substituting now  \ref{11-theta-theta-1}, into \ref{theta_1}, we obtain after some algebra
\begin{eqnarray}
\label{11-theta-theta-3'}
\langle{0}|{T^\theta_\theta}|{0}\rangle&=&{\frac{\eta }{4\pi^2\beta^3 \alpha^{4} r}\int_0^\infty{dy}\;y\;e^{y\left(1-r^2/\eta^2\right)}\text{Re}\left[K_{1/2-im\alpha}(y)\right]} \sum_{j=1/2}^{\infty}(2j+1)^2\left(I_{\frac{j+1/2}{\beta}-1/2}(yr^2/\eta^2)-I_{\frac{j+1/2}{\beta}+1/2}(yr^2/\eta^2)\right)\nonumber\\
\end{eqnarray}
Note that the angular symmetry of the our background \ref{metric2} trivially guarantees that $\langle{0}|{T^\phi_\phi}|{0}\rangle=\langle{0}|{T^\theta_\theta}|{0}\rangle$.\\ 

Finally, using \ref{psi7p} and \ref{basis} into \ref{T2}, it can be easily seen that $\langle 0 |T_{\mu}{}^{\nu}|0\rangle =0$, for all $\mu \neq \nu$, leaving us only with the diagonal components for the vacuum expectation value.

\section{Pure global monopole contribution to $\langle 0| T_{\mu}{}^{\nu}|0\rangle$} \label{s5}
Now as of \ref{s3}, we wish to extract below the pure global monopole contribution from the expressions \ref{energy-density3}, \ref{rr-emt1}, \ref{11-theta-theta-3'}.  We write in analogy of \ref{conds4}
\begin{eqnarray}
\label{pure-gm-1}
\langle{0}|{T^\nu_\mu}|{0}\rangle_{\rm gm}&=&\langle{0}|{T^\nu_\mu}|{0}\rangle-\langle{0}|{T^\nu_\mu}|{0}\rangle_{\rm dS},
\end{eqnarray}
The derivation of the pure de Sitter contribution, $\langle{0}|{T_\mu{}^\nu}|{0}\rangle_{\rm dS}$ (corresponding to $\beta=1$ in  \ref{energy-density3}, \ref{rr-emt1}, \ref{11-theta-theta-3'}) is very briefly sketched in \ref{C}, once again for a check of consistency of our mode functions. Note that $\langle{0}|{T^\nu_\mu}|{0}\rangle_{\rm dS}$ has been computed in numerous places earlier, 
e.g. in~\cite{BezerradeMello:2010ci} in the context of de Sitter cosmic strings. Owing to the maximal symmetry of the de Sitter spacetime, each component of this vacuum expectation value is the same constant, \ref{emt-ds-ren-pure}.

Subtracting now the first equation of \ref{ds-emt-tt} respectively from \ref{energy-density3}, \ref{rr-emt1}, \ref{11-theta-theta-3'}, we have the corresponding pure global monopole contributions,
\begin{eqnarray}
\langle{0}|T_0{}^0|{0}\rangle_{\rm gm}&&=\frac{\eta }{2\pi^2\alpha^{4}{r}}\int_0^\infty{dy}\;y\;e^{y\left(1-r^2/\eta^2\right)}\text{Re}[K_{1/2-im\alpha}(y)]\sum_{j=1/2}^{\infty}\frac{(j+1/2)}{\beta^2}\nonumber\\ &&\times\Bigg[\left(I_{\frac{j+1/2}{\beta}+1/2}(yr^2/\eta^2)-I_{{j+1/2}+1/2}(yr^2/\eta^2)\right)+\left(I_{\frac{j+1/2}{\beta}-1/2}(yr^2/\eta^2)-I_{{j+1/2}-1/2}(yr^2/\eta^2)\right)\Bigg],\nonumber\\
\langle{0}|{T_r{}^r}|{0}\rangle_{\rm gm}&&=\langle{0}|T_0{}^0|{0}\rangle_{\rm gm},\nonumber\\
\langle{0}|{T_\theta{}^\theta}|{0}\rangle_{\rm gm}&&=\langle{0}|{T_{\phi}{}^\phi}|{0}\rangle_{\rm gm}=\frac{\eta }{2\pi^2\alpha^{4}{r}}\int_0^\infty{dy}\;y\;e^{y\left(1-r^2/\eta^2\right)}\text{Re}[K_{1/2-im\alpha}(y)]\sum_{j=1/2}^{\infty}\frac{(j+1/2)^2}{\beta^3}\nonumber\\
&&\times\left[\left(I_{\frac{j+1/2}{\beta}-1/2}(yr^2/\eta^2)-I_{{j+1/2}-1/2}(yr^2/\eta^2)\right)-\left(I_{\frac{j+1/2}{\beta}+1/2}(yr^2/\eta^2)-I_{{j+1/2}+1/2}(yr^2/\eta^2)\right)\right]\nonumber\\
\end{eqnarray}
Applying now the Abel-Planna summation formula~\ref{abel-planna1} into the above equations, we get
\begin{eqnarray}
\label{final-gm}
\langle{0}|T_0{}^0|{0}\rangle_{\rm gm}&=&\langle{0}|{T_r{}^r}|{0}\rangle_{\rm gm}=\frac{2\eta }{\pi^3\alpha^{4}{r}}\int_{0}^\infty{du u g(\beta,u)}\int_0^\infty{dy} y e^{y\left(1-r^2/\eta^2\right)}\text{Re}[K_{1/2-im\alpha}(y)]\text{Im}\left[K_{1/2-iu}\left(yr^2/\eta^2\right)\right],\nonumber\\
\langle{0}|{T_\theta{}^\theta}|{0}\rangle_{\rm gm}&=&\langle{0}|{T_\phi{}^\phi}|{0}\rangle_{\rm gm}=\frac{2\eta}{\pi^3\alpha^{4}{r}}\int_{0}^\infty{du u^2 g(\beta,u)}\int_0^\infty{dy}\;y\;e^{y\left(1-r^2/\eta^2\right)}\text{Re}[K_{1/2-im\alpha}(y)]\text{Re}\left[K_{1/2-iu}\left(yr^2/\eta^2\right)\right]\nonumber\\
\end{eqnarray}
where the function $g(\beta,u )$ is defined in  \ref{g}. Also, in deriving  $\langle{0}|{T_\theta{}^\theta}|{0}\rangle_{\rm gm}$  (or $\langle{0}|{T_\phi{}^\phi}|{0}\rangle_{\rm gm}$), we have used,
$$\sum_{j=1/2}\frac{1}{\beta}\left(\frac{j+1/2}{\beta}\right)^2\left(\left(I_{\frac{j+1/2}{\beta}-1/2}-I_{j+1/2-1/2}\right)-\left(I_{\frac{j+1/2}{\beta}+1/2}-I_{j+1/2+1/2}\right)\right)=\frac{4}{\pi}\int_{0}^{\infty}du\;u^2 g(\beta,u)\text{Re}[K_{1/2-iu}(yr^2/\eta^2)]$$

Let us now take the Minkowski limit of the above expressions. This should correspond to $t/\alpha \to 0$ in the metric, where $t$ is the cosmological time. This means the conformal time $\eta \approx -\alpha$ in this limit. Thus in \ref{final-gm}, defining a variable $z=y r^2/ \alpha^{2}$, we have for a fixed $z$-value and large $\alpha$~\cite{Abra64},
\begin{eqnarray}
\label{mink-lim1}
\text{Re}[K_{1/2-i;m\alpha}(z\alpha^2)]\approx \sqrt{\frac{{\pi}}{2z \alpha^2}}e^{-\frac{m^2}{2z}-z\alpha^2}
\end{eqnarray}
Substituting the above into \ref{final-gm}, we have
\begin{eqnarray}
\label{Minkgm}
\langle{0}|T_0{}^0|{0}\rangle_{\rm gm}^{\rm (M)}&=&\langle{0}|T_r{}^r|{0}\rangle_{\rm gm}^{\rm (M)}=\frac{2^{1/2}}{\pi^{5/2}{r^4}}\int_{0}^\infty{du\;u\;g(\beta,u)}\int_0^\infty{dz}\;z^{1/2}e^{-\frac{m^2r^2}{2z}-z}\text{Im}\left[K_{1/2-iu}\left(z\right)\right]\nonumber\\
\langle{0}|T_{\theta}{}^\theta |{0}\rangle_{\rm gm}^{\rm (M)}&=&\langle{0}|T_{\phi}{}^\phi |{0}\rangle_{\rm gm}^{\rm (M)}=\frac{2^{1/2}}{\pi^{5/2}{r^4}}\int_{0}^\infty{du\;u^2\;g(\beta,u)}\int_0^\infty{dz}\;z^{1/2}e^{-\frac{m^2r^2}{2z}-z}\text{Re}\left[K_{1/2-iu}\left(z\right)\right]
\end{eqnarray}
Setting $m^2=0$ above, let us compute the trace of $\langle{0}|T_\mu{}^\nu|{0}\rangle_{\rm gm}^{\rm (M)}$. We have 
\begin{eqnarray}
\label{renorm0}
\langle{0}|T_\mu{}^\mu|{0}\rangle_{\rm gm}^{(M)}\big\vert_{m^2=0}&=&\frac{2^{3/2}}{\pi^{5/2}{r^4}}\int_{0}^\infty{du\;u\;g(\beta,u)}\int_0^\infty{dz}\;z^{1/2}e^{-z}\left(\text{Im}\left[K_{1/2-iu}\left(z\right)\right]+{u}\text{Re}\left[K_{1/2-iu}\left(z\right)\right]\right) \qquad 
\end{eqnarray}
We now separate the right hand side of  \ref{identity1}  into real and imaginary parts and substitute into the above integral. It is easy to see that both the $z$-integrals vanish.  Thus 
the trace anomaly induced by the global monopole in the flat spacetime for fermionic field is vanishing.\\

Likewise, we can show that the trace anomaly induced by the global monopole in the de Sitter spacetime is vanishing, too.
Setting  $m=0$ in \ref{final-gm}, we have
\begin{eqnarray}
\label{trace-anomaly}
\langle{0}|T_\mu{}^\mu|{0}\rangle_{\rm gm}\big\vert_{m=0}&=&\frac{2^{3/2}}{\pi^{5/2}\left(\alpha{r}/\eta\right)^4}\int_{0}^\infty{du\;u\;g(\beta,u)}\int_0^\infty{dx}\;x^{1/2}\;e^{-x}\left(\text{Im}\left[K_{1/2-iu}\left(x\right)\right]+{u}\text{Re}\left[K_{1/2-iu}\left(x\right)\right]\right) \qquad 
\end{eqnarray}
where we have used  $K_{1/2}=\sqrt{\pi}e^{-y}/2y$~\cite{BezerradeMello:2010ci}, and the variable transformation $x=yr^2/\eta^2$ to arrive at the above expression. The above integral is formally similar to \ref{renorm0} and hence is vanishing. 

The vanishing of the above trace anomalies can also be seen from the vanishing of the condensate $\langle 0| \overline{\Psi} \Psi | 0\rangle_{\rm gm} $ for $m=0$, as discussed in \ref{s3}. From \ref{emt1} and the Dirac equation, the trace anomaly for the fermionic field is given by 
\be
\langle 0|T_{\mu}{}^{\mu}| 0\rangle\big\vert_{m=0}= \lim_{m\to 0}m \langle 0| \overline{\Psi} \Psi | 0\rangle
\label{add}
\ee
Thus it is clear that the trace anomaly induced by the global monopole in the de Sitter spacetime (and hence in the Minkowski spacetime) should be vanishing. This also shows that the decompositions of \ref{conds4} and \ref{pure-gm-1} are consistent with each other.

It can also be easily checked that the global monopole contribution to the energy momentum tensor \ref{final-gm}, obeys the  conservation  equation, $\nabla_{\mu}{\langle{T^\mu{}_\nu}\rangle_{\rm gm}}=0$. (The pure de Sitter part \ref{emt-ds-ren-pure}, satisfies the same trivially). In the background \ref{metric2}, we have two independent components of $\nabla_{\mu}{\langle{T^\mu{}_\nu}\rangle_{\rm gm}}$, 
\begin{eqnarray}
\label{cov1}
&&\partial_\eta{\langle{T^0_0}\rangle_{\rm gm}}+\frac{1}{\eta}\left({\langle{T^r_r}\rangle_{\rm gm}}+2{\langle{T^\theta_\theta}\rangle_{gm}}-3{\langle{T^0_0}\rangle_{\rm gm}}\right),\quad {\rm and} \quad  \partial_r{\langle{T^r_r}\rangle_{\rm gm}}+\frac{2}{r}\left({\langle{T^r_r}\rangle_{\rm gm}}-{\langle{T^\theta_\theta}\rangle_{\rm gm}}\right)
\end{eqnarray}
Substituting the components from \ref{final-gm}, it is easy to check that both the above expressions vanish.\\

\begin{figure}[ht]
\begin{center}
    \includegraphics[scale=0.55]{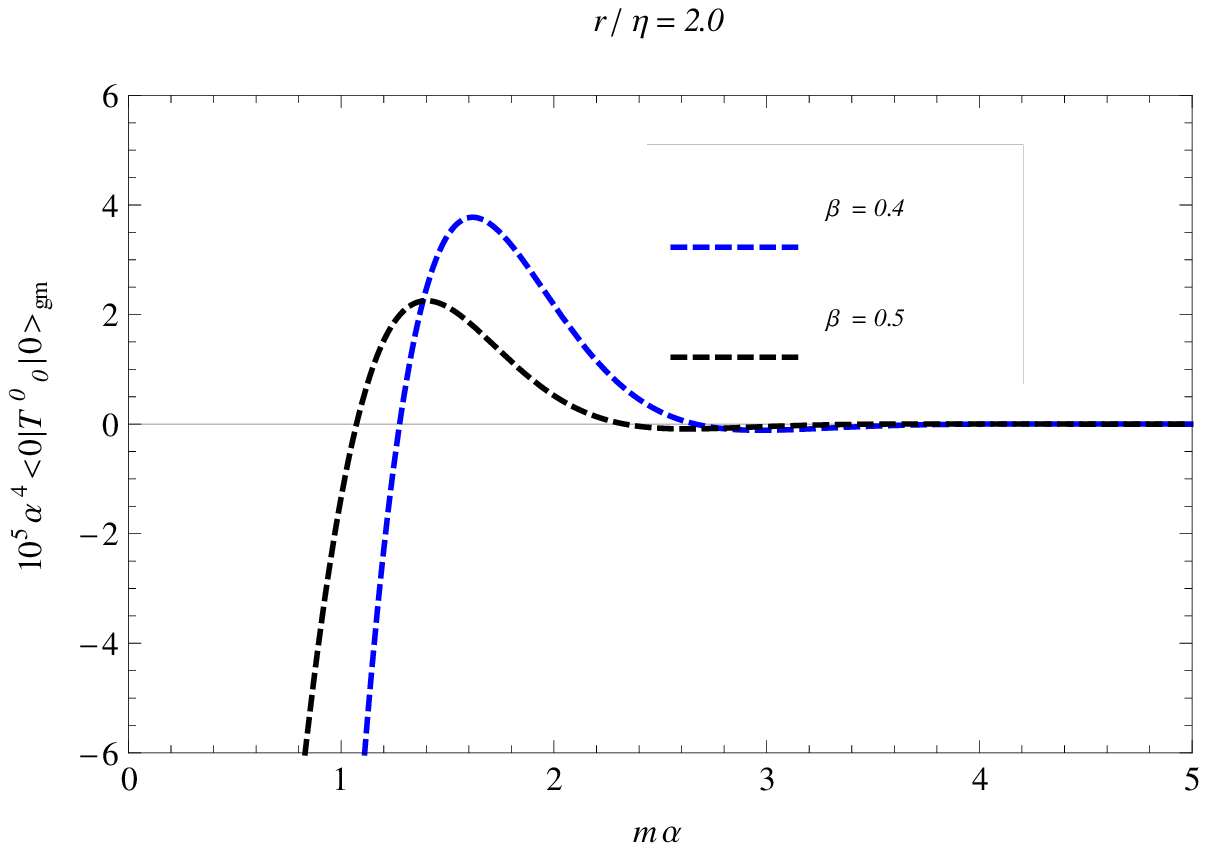}
    \includegraphics[scale=0.55]{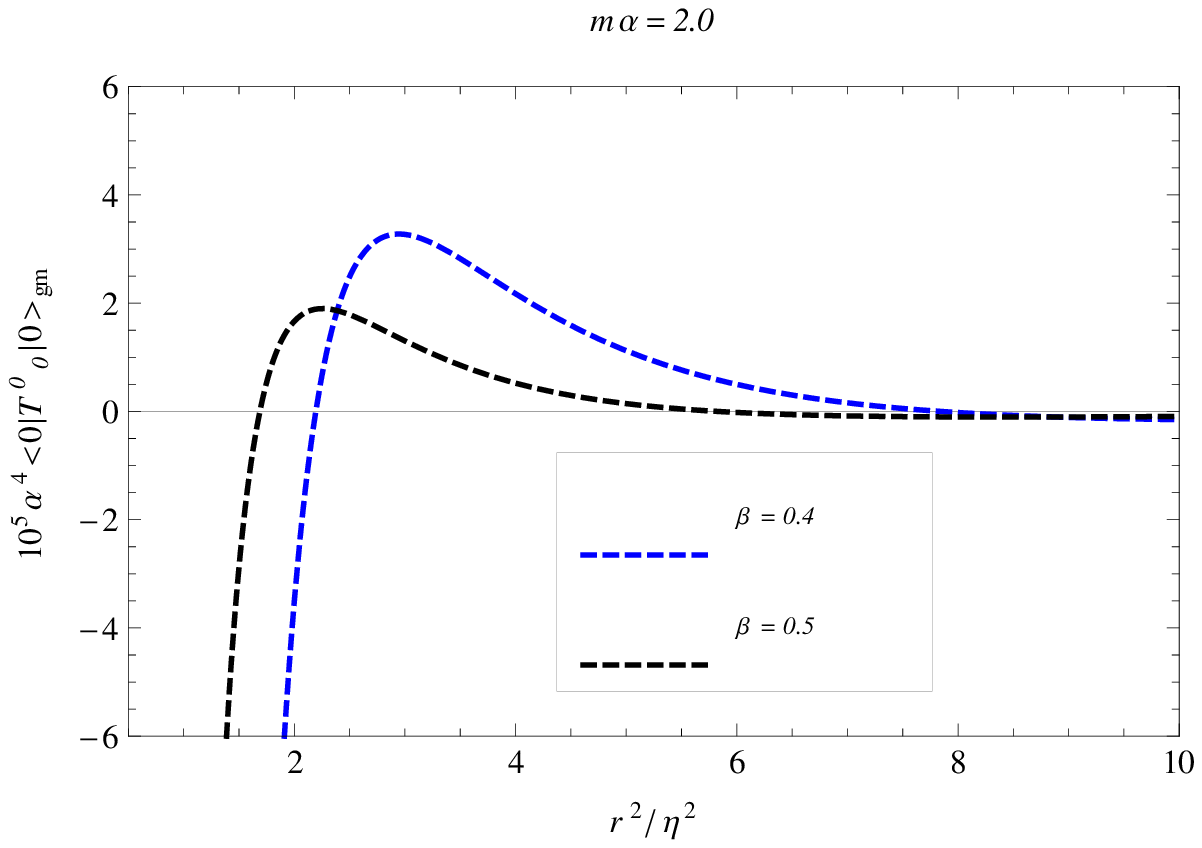}
\end{center}
\caption{Variation of the vacuum expectation values of energy density or the radial pressure of the global monopole induced part (the first of \ref{final-gm}) as a function of the dimensionless mass parameter $m\alpha$ (left) and the dimensionless  radial distance (right). The $m\to0$ limit does not diverge, but touch the $y$-axis at some high negative value. On the other hand, the $r/|\eta| \to 0$ limit is divergent, owing to the curvature singularity located there due to the global monopole. See main text for detail.}
\label{energy}
\end{figure}
\begin{figure}
    \includegraphics[scale=0.65]{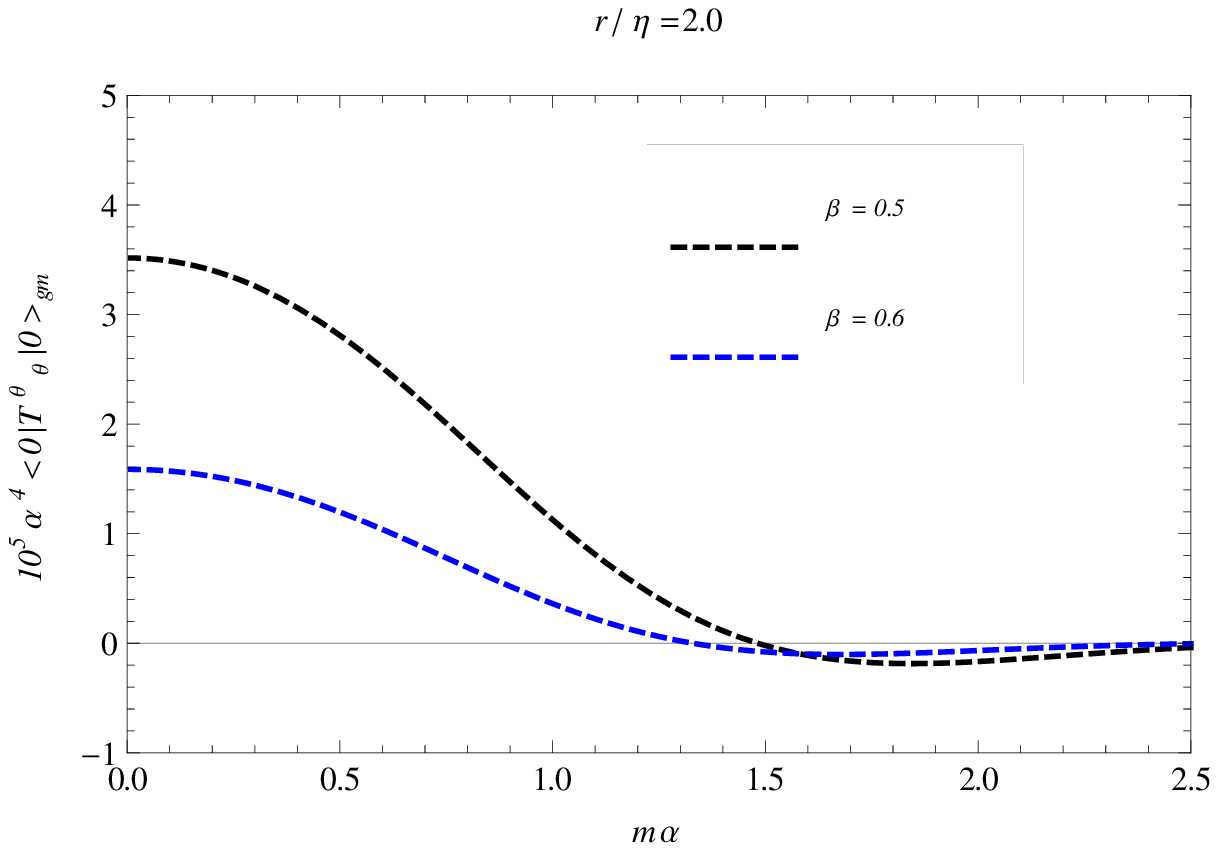}\hspace{0.2mm}
    \includegraphics[scale=0.65]{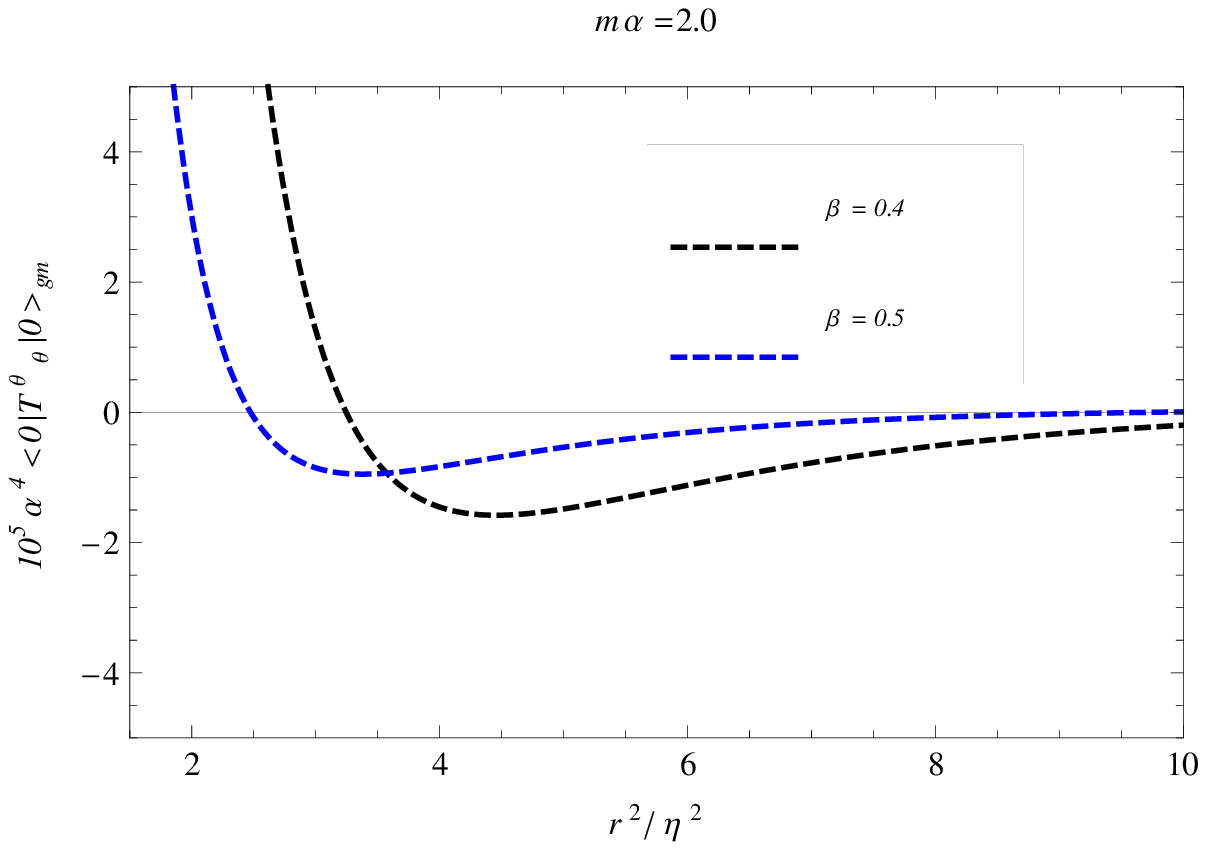}
\caption{The variation of the vacuum expectation values of the polar or azimuthal stresses induced by the global monopole,  \ref{final-gm}, as a function of the dimensionless mass parameter $m\alpha$ (left) and the radial distance $r$/$\eta$ (right). See main text for detail.}
\label{angular}
\end{figure}

As of \ref{s3}, let us now look into the two special cases of \ref{final-gm}, i.e. small and large proper distances from the  monopole. Using for the real part of the modified Bessel function~\cite{Abra64} for small argument,
$$\text{Re}\left[{K_{1/2-im\alpha}}(y)\right]\approx{\sqrt{\frac{\pi}{2y}}}e^{-y}\left(1-\frac{m^2\alpha^2}{2y}\right)$$
Substituting the above into the first of \ref{final-gm}, we have
\begin{eqnarray}
\label{tt-gm}
\langle{0}|T^0_0|{0}\rangle_{\rm gm}\big\vert_{r/|\eta| \to 0}&&=\frac{2^{1/2}\eta }{\pi^{5/2}\alpha^{4}{r}}\int_{0}^\infty{du\;u\;g(\beta,u)}\int_0^\infty{dy}\;y^{1/2}\;e^{-yr^2/\eta^2}\left(1-\frac{m^2\alpha^2}{2y}\right)\text{Im}\left[K_{1/2-iu}\left(yr^2/\eta^2\right)\right]\nonumber\\
&&=\langle{0}|T^r_r|{0}\rangle_{\rm gm}\big\vert_{r/|\eta| \to 0}
\end{eqnarray}
which can be evaluated exactly using \ref{identity1}, \ref{express1}
\begin{eqnarray}
\label{tt1-gm}
\langle{0}|T^0_0|{0}\rangle_{\rm gm}\big\vert_{r/|\eta| \to 0}&&=-\frac{1}{72\pi}\bigg(\frac{7}{40\left( r\alpha/\eta\right)^{4}}\left(\frac{1}{\beta^4}-1\right)+\frac{3}{4\pi^3 \left(\alpha r/\eta\right)^4}\sum_{n=0}^\infty (-1)^{n} \big({\zeta(4,1+(n+1)\beta)}-{\zeta(4,1+(n+1))}\big)\nonumber\\
&&-\frac{\pi m^2}{3\left(r\alpha/\eta\right)^2}\left(\frac{1}{\beta^2}-1\right)+\frac{9m^2}{2\pi \left(\alpha r/\eta\right)^2}\sum_{n=0}^\infty (-1)^{n} \left[{\zeta(2,1+(n+1)\beta)}-{\zeta(2,1+(n+1))}\right]\bigg)\nonumber\\&&=\langle{0}|T^r_r|{0}\rangle_{\rm gm}\big\vert_{r/|\eta| \to 0}
\end{eqnarray}
where the multiple $\zeta$ function is defined below \ref{express1}. Likewise we have
\begin{eqnarray}
\label{radial_1-gm}
\langle{0}|T^\theta_\theta|{0}\rangle_{\rm gm}\big\vert_{r/|\eta| \to 0}&=&\langle{0}|T^\phi_\phi|{0}\rangle_{\rm gm}=\frac{1}{72\pi}\bigg(\frac{7}{40\left( r\alpha/\eta\right)^{4}}\left(\frac{1}{\beta^4}-1\right)+\frac{3}{4\pi^3\left(\alpha r/\eta\right)^4}\sum_{n=0}^\infty (-1)^{n} \big({\zeta(4,1+(n+1)\beta)}\nonumber\\
&-&{\zeta(4,1+(n+1))}\big)-\frac{9m^2\zeta(3)}{\pi^2\left(r\alpha/\eta\right)^2}\left(\frac{1}{\beta^3}-1\right)+\frac{18m^2}{\pi^2 \left(\alpha r/\eta\right)^2}\sum_{n=0}^\infty (-1)^{n} \big({\zeta(3,1+(n+1)\beta)}\nonumber\\
&-&{\zeta(3,1+(n+1))}\big)\bigg)
\end{eqnarray}
Note that  all the components diverge at the location of the monopole $(r=0)$, due the curvature singularity present there.

As earlier, the next special case is to take small value of $y$ in \ref{final-gm}, corresponding to large proper radial distances from the global monopole. Following procedure similar to that of described at the end of \ref{s3}, we find
\begin{eqnarray}
\label{final-gm0}
\langle{0}|T^0_0|{0}\rangle_{\rm gm}\big\vert_{r/|\eta| \to \infty}&=&\frac{f(\beta,m\alpha)}{\sqrt{2}\pi^{5/2}\left(r\alpha/|\eta|\right)^{4}}\cos\left(2m\alpha\text{ln}\left(2r/|\eta|\right)-\varphi_0\right)=\langle{0}|{T^r_r}|{0}\rangle_{\rm gm}\big\vert_{r/|\eta| \to \infty}\nonumber\\
\langle{0}|{T^\theta_\theta}|{0}\rangle_{\rm gm}\big\vert_{r/|\eta| \to \infty}&=&\langle{0}|{T^\phi_\phi}|{0}\rangle_{\rm gm}\big\vert_{r/|\eta| \to \infty}=\frac{f(\beta,m\alpha)}{\sqrt{2}\pi^{5/2}\left(r\alpha/|\eta|\right)^{4}}(1+m^2\alpha^2)^{1/2}\,\sin\left(2m\alpha\text{ln}\left(2r/|\eta|\right)-\varphi_0+\phi_1\right)\qquad
\end{eqnarray}
where $\phi_1=\cot^{-1}m\alpha$, and the function $f(\beta,m\alpha)$ and the phase factor $\phi_0$ are defined in \ref{B-function}.  Note that both the short distance divergence (\ref{tt1-gm}, \ref{radial_1-gm}) and the 
rapid fall-off with oscillation (\ref{final-gm0})  are qualitatively similar to that of the vacuum condensate discussed in \ref{s3}. This is also similar  to the case  of the de Sitter cosmic string~\cite{BezerradeMello:2010ci}.  However, this is not exactly the case for a scalar field located in the de Sitter global monopole with a finite core background~\cite{BezerradeMello:2009ju}. For this case we have off-diagonal components of the energy-momentum tensor as well as a milder divergence at short distance. \\

Finally, we have plotted the variations of \ref{final-gm} in \ref{energy} and \ref{angular}, 
with respect to the dimensionless mass parameter $m\alpha$, as well as the dimensionless radial distance $r/|\eta|$, with different values of the defect parameters. Note that for heavily massive cases, $m\alpha \gg 1$,  the expectation values are highly suppressed.

\section{Conclusions}\label{s6}
In this work we have computed the vacuum expectation values of $\overline {\Psi}\Psi$ and the energy-momentum tensor of the Dirac fermions in the de Sitter spacetime endowed with a pointlike global monopole defect.  We have extracted the pure global monopole contributions for various quantities, \ref{gmconds1}, \ref{final-gm}, by subtracting the pure de Sitter part from the full contribution.  These results contain radial dependences as a consequence of the breaking of the translational symmetry  of the de Sitter spacetime due to the presence of the global monopole located at $r=0$. The general results \ref{gmconds1}, \ref{final-gm}, could not further be simplified analytically and hence was plotted in~\ref{f1}, \ref{energy}, \ref{angular}. However, in the special scenarios when we are sufficiently close to, or far away from the monopole are also analysed analytically in \ref{s3}, \ref{s5}.

Note that all the results we have found diverge at the the location of the monopole due to the curvature singularity introduced by it and fall of sufficiently rapidly for large proper radial distances. We also note that the vanishing of the fermionic condensate, \ref{s3}, for the global monopole indicates vanishing of the trace anomaly due to itself as well, which has been confirmed in \ref{s5}. This also shows that the decompositions of \ref{conds4} and \ref{pure-gm-1} are consistent with each other.
 However, we must also note here an essential caveat regarding what we mean by the `global monopole induced' part in all the above results. Precisely, the expressions for anomaly are universal, given by curvature invariants, e.g.~\cite{Parker:2009uva}. The curvature of \ref{metric1} should also get contribution from the defect $\beta <1$, even for $\alpha\to \infty$. Indeed, by this method we directly have the anomaly
 $$\langle T_{\mu}{}^{\mu}\rangle= \frac{4b(1-\beta^2)^2}{3\beta^4(\alpha r/\eta)^4}+\frac{8b'\left(r^4(9-8(1-\beta^2))-(\eta^2-r^2)^2(1-\beta^2) \right)}{3\eta^4 \beta^2 (\alpha r/\eta)^4}+\frac{4b''(1-\beta^2)(\eta^2+r^2)}{\eta^2\beta^2(\alpha r/\eta)^4}$$
 where $b=1/320\pi^2$ and $b'=-11/5760\pi^2$. The coefficient $b''$ is not fixed, and can be changed by adding a term proportional to $R^2$ in the gravitational action. Assuming $\beta$ to be close to unity, we may identify the terms dependent on $(1-\beta^2)$ or its powers in the above expression to be the part induced by the global monopole, it is clear that no choice of $b''$ can make the anomaly vanishing, leading to a contradiction to what we have found above using  the Dirac modes, \ref{psi7p}, even though  \ref{energy}, \ref{angular} show that an individual massless $\langle T_{\mu}{}^{\nu}\rangle$ is indeed dependent on the defect, $\beta$. It seems similar mismatch is present for a de Sitter cosmic string as well~\cite{BezerradeMello:2010ci}.  Keeping in mind that the anomaly is essentially an ultraviolet phenomenon,  this ambiguity could probably be related to the particular regularisation scheme we have adopted here, i.e. subtracting the de Sitter contribution from the full expression, \ref{conds4}, \ref{pure-gm-1}. The standard  regularisation techniques in curved spacetime on the other hand, apart from the dimensional regularisation, either use the so called adiabatic subtraction, or subtract the flat spacetime contribution from the Hadamard expansion of the Green function, obtained via the Riemann normal coordinates~\cite{Parker:2009uva}. Thus possibly we are loosing some essential curvature contribution, e.g. via equations like \ref{abel-planna3}, \ref{g}. It is well known that the results of quantum field theory in curved spacetimes can be regularisation dependent. Perhaps one should compute the fermion Green function using the modes of \ref{psi7p} and use some other regularisation technique to see if it reproduces the above mentioned non-vanishing expression of anomaly due to the global monopole. We reserve this for a future work. 

There can be further some directions towards which our present study can be extended.  For example,  the orthonormal Dirac modes found here  may be used to compute cosmological correlation functions. In particular as we stated above, we may use them to find out the Green or the Wightman functions, which can be used in perturbative calculations  in the presence of some  interactions. 
It will further be interesting to investigate the vacuum expectation value of the conserved current and in particular the chiral anomaly in this background.  We hope to return to these issues in our future works.

\bigskip
\section*{Acknowledgement}
The research of M.S.A was supported by the ISIRD grant 9-252/2016/IITRPR/708. The research of SB was partially supported by the ISIRD grant 9-289/2017/IITRPR/704. We would like to acknowledge anonymous referee for useful  critical comments on an earlier version of this manuscript. The research of M. S. A. is also supported by the National Postdoctoral Fellowship of the Science and Engineering Research Board (SERB), Department of Science and Technology (DST), Government of India, file No., PDF/2021/003491.
\bigskip
\appendix
\labelformat{section}{Appendix #1} 
\section{Detail for \ref{s2}}\label{A}

In this Appendix we shall provide the computational details for \ref{s2}. The Dirac equation in curved spacetime reads,
\begin{eqnarray}
\label{dirac1}
\left(i \gamma^{\mu}\nabla_\mu-m\right)\Psi(x)=0,
\end{eqnarray}
where $\nabla_\mu \equiv \partial_\mu+\Gamma_\mu$ is the spin covariant derivative, and $\Gamma_\mu$'s are the spin connection matrices, 
\begin{eqnarray}
\label{gamma1}
\Gamma_\mu=\frac{1}{4}\gamma^{(a)}\gamma^{(b)}e^{\nu}_{(a)} \nabla_{\mu}e_{(b)\nu}
\end{eqnarray}
where the Latin indices within parenthesis stand for the local Lorentz frame and  $\gamma^\mu=e^{\mu}_{(a)}\gamma^{(a)}$, where  $e^{\mu}_{(a)}$'s are the tetrads.

 Since the  metric \ref{metric2} is just conformally related to the flat metric with a global monopole, it is clear that the angular part of our modes will be the same as that of the latter, can be seen in e.g.~\cite{Saharian:2005br}. 
 
 For \ref{metric2}, we choose the tetrads to be
\begin{eqnarray}
\label{basis}
e^{\mu}_{(a)}=\frac{\eta}{\alpha}\left({\begin{array}{cccc}
   1 & 0 & 0 &0\\
  0 & {\sin\theta\cos\phi}  &\frac{\cos\theta\cos\phi}{\beta r}& -\frac{\sin\phi}{\beta r\sin\theta}\\
  0 & {\sin\theta\sin\phi} 
  &\frac{\cos\theta\sin\phi}{\beta r}& 
  \frac{\cos\phi}{\beta r\sin\theta}\\
   0 & {\cos\theta}
   &-\frac{\sin\theta}{\beta r}& 0\\
  \end{array} } \right),
\end{eqnarray} 
where the rows of the matrix are specified by the local Lorentz index $a$ and the columns by the spacetime index $\mu$. From \ref{gamma1}, \ref{basis}, we find the non-vanishing  spin connection matrices
\begin{eqnarray}
\label{gamma2'}
\Gamma_1=-\frac{g_{rr}}{2\eta}\gamma^{1}\gamma^{0},\qquad \Gamma_2=-\frac{g_{\theta\theta}}{2\eta}\gamma^{2}\gamma^{0}+\frac{g_{\theta\theta}}{2r}\left(1-\frac{1}{\beta}\right)\gamma^{2}\gamma^{1},\qquad 
\Gamma_3=-\frac{g_{\phi\phi}}{2\eta}\gamma^{3}\gamma^{0}+\frac{g_{\phi\phi}}{2r}\left(1-\frac{1}{\beta}\right)\gamma^{3}\gamma^{1}
\end{eqnarray}
which yields,
\begin{eqnarray}
\label{gamma3}
\gamma^\mu\Gamma_\mu=-\frac{3}{2\alpha}\gamma^{(0)}+\frac{\beta-1}{\beta r}\gamma^{(1)}
\end{eqnarray}
We take the representation of the $\gamma$-matrices,
\begin{eqnarray}
\label{gamma2}
\gamma^{(0)}=\left( {\begin{array}{cccc}
   I & 0\\
  0 & -I\\
  \end{array} } \right),\qquad \gamma^{(i)}=\left( {\begin{array}{cccc}
   0 & \sigma^i\\
  -\sigma^i & 0\\
  \end{array} } \right),
  \end{eqnarray} 
where $\sigma^i$'s are the  Pauli matrices. 
Putting now things in together, the Dirac equation \ref{dirac1} takes the form,
\begin{eqnarray}
\label{dirac1'}
&&\left[\left(\gamma^{(0)}\partial_\eta-\frac{3}{2\eta}\gamma^{(0)}+\frac{i m\alpha}{\eta}\right)+\vec{\gamma}\cdot\hat{r}\left(\partial_r+\frac{\beta-1}{\beta r}\right)+\frac{1}{\beta r}\left(\vec{\gamma}\cdot\hat{\theta}\partial_\theta+\frac{\vec{\gamma}\cdot\hat{\phi}}{\sin\theta}\partial_\phi\right)\right]\Psi=0,
\end{eqnarray}
where $\hat{r},\;\hat{\theta},\;\text{and}\;\hat{\phi}$ are the unit vectors in spherical polar coordinates. We now take
\begin{eqnarray}
\label{Psi1}
\Psi=\left( {\begin{array}{c}
  \Psi_1\\
 \Psi_2\\
  \end{array} } \right)
\end{eqnarray}
where $\Psi_1$ and $\Psi_2$ are both two component spinors. Using now \ref{gamma2}, we have from \ref{dirac1'},
\begin{eqnarray}
\label{dirac2}
\left(\partial_\eta-\frac{1}{\eta}\left(\frac{3}{2}-\frac{i m\alpha}{\eta}\right)\right)\Psi_1+\left(\vec{\sigma}\cdot\hat{r}\left(\partial_r+\frac{\beta-1}{\beta r}\right)+\frac{1}{\beta r}\left(\vec{\sigma}\cdot\hat{\theta}\partial_\theta+\frac{\vec{\sigma}\cdot\hat{\phi}}{\sin\theta}\partial_\phi\right)\right)\Psi_2 &=&0,\nonumber\\
\left(\partial_\eta-\frac{1}{\eta}\left(\frac{3}{2}+\frac{i m\alpha}{\eta}\right)\right)\Psi_2+\left(\vec{\sigma}\cdot\hat{r}\left(\partial_r+\frac{\beta-1}{\beta r}\right)+\frac{1}{\beta r}\left(\vec{\sigma}\cdot\hat{\theta}\partial_\theta+\frac{\vec{\sigma}\cdot\hat{\phi}}{\sin\theta}\partial_\phi\right)\right)\Psi_1 &=&0
\end{eqnarray}
Using now (e.g.~\cite{Berestetsky:1982aq})
\begin{eqnarray}
\label{sigma6}
\left(\vec{\sigma}\cdot\hat{\theta}\partial_\theta+\frac{\vec{\sigma}\cdot\hat{\phi}}{\sin\theta}\partial_\phi\right)\equiv -\vec{\sigma}\cdot\hat{r}\left(\hat{K}-1\right),\qquad {\rm with} \qquad \hat{K}=\vec{\sigma}\cdot\vec{L}+1,
\end{eqnarray}
where $\vec{L}$ is the standard orbital angular momentum operator,  \ref{dirac2} can be rewritten as,
\begin{eqnarray}
\label{dirac3}
\left(\partial_\eta-\frac{1}{\eta}\left(\frac{3}{2}-\frac{i m\alpha}{\eta}\right)\right)\Psi_1+\vec{\sigma}\cdot\hat{r}\left(\partial_r+\frac{1}{r}-\frac{\hat{K}}{\beta r}\right)\Psi_2 &=&0,\nonumber\\
\left(\partial_\eta-\frac{1}{\eta}\left(\frac{3}{2}+\frac{im\alpha}{\eta}\right)\right)\Psi_2+\vec{\sigma}\cdot\hat{r}\left(\partial_r+\frac{1}{r}-\frac{\hat{K}}{\beta r}\right)\Psi_1 &=&0
\end{eqnarray}
We now choose the ansatz for the variable separation 
\begin{eqnarray}
\label{psi2}
\Psi_1=f^{\sigma}(r,\eta)\Omega_{jl_\sigma m}(\theta, \phi),\qquad \Psi_2=(-1)^\sigma g^{\sigma}(r,\eta)\Omega_{jl_\sigma' m}(\theta, \phi) \qquad (\sigma =0,\,1,\,\,\,{\rm no~sum~on~}\sigma)
\end{eqnarray}
where 
\begin{eqnarray}
\label{sigma7}
l_\sigma=j-\frac{(-1)^\sigma}{2},\;l_\sigma'=j+\frac{(-1)^\sigma}{2},
\end{eqnarray}
and $\Omega_{jl_\sigma m}$ are the standard  spin-1/2 spherical  harmonics 
$$\Omega_{l-1/2,l,m}\left(\theta,\phi\right)=\left( {\begin{array}{cc}
    C^{+}_{l{m}}Y_{l,m-1/2}\left(\theta,\phi\right)\\
     C^{-}_{l{m}}Y_{l,m+1/2}\left(\theta,\phi\right) \\
  \end{array} } \right),\;\;\;\Omega_{l+1/2,l,m}\left(\theta,\phi\right)=\left( {\begin{array}{cc}
  -C^{-}_{l{m}}Y_{l,m-1/2}\left(\theta,\phi\right) \\
    C^{+}_{l{m}}Y_{l,m+1/2}\left(\theta,\phi\right)\\
  \end{array} } \right)$$
  where
  $$C^{\pm}_{l{m}}=\sqrt{\frac{{l\pm{m}+1/2}}{2l+1}}$$
 $j(j+1)$ is the eigenvalue of the total angular momentum ($J^2$), $j=1/2,3/2,...$ and $m=-j,...,j$ are its projections. 
$\Omega_{jl_\sigma m}$'s are the simultaneous eigenfunctions of the operators $L^2,\;S^2,\;J^2,\;J_z$ and $\hat{K}$. In particular 
(e.g.~\cite{Berestetsky:1982aq, Bhattacharya:2019zno, Saharian:2005br} and references therein),
\begin{eqnarray}
\label{opertaor1'}
\hat{K}\Omega_{jl_\sigma m}=-k_\sigma\Omega_{jl_\sigma m},\;\;k_\sigma=-{(-1)^\sigma\left(j+1/2\right)}.
\end{eqnarray}
Putting $j=l+1/2$, we have
\begin{eqnarray}
\label{opertaor1}
(\vec{\sigma}\cdot\hat{r})\Omega_{l+1/2,l, m}=-\Omega_{(l+1)-1/2,l+1, m},\;\;\hat{K}\Omega_{l+1/2,l, m}=(l+1)\Omega_{l+1/2,l, m}\nonumber\\
(\vec{\sigma}\cdot\hat{r})\Omega_{(l+1)-1/2,l+1, m}=-\Omega_{l+1/2,l, m},\;\;\hat{K}\Omega_{(l+1)-1/2,l+1, m}=-(l+1)\Omega_{(l+1)-1/2,l+1, m}
\end{eqnarray}
Using now
\begin{eqnarray}
\label{Omega}
\Omega_{jl_\sigma' m}=i^{l_\sigma-l_\sigma'}(\hat{r}\cdot\vec{\sigma})\Omega_{jl_\sigma m},
\end{eqnarray}
\ref{dirac3} can be recast as
\begin{eqnarray}
\label{dirac4}
\left(\frac{\partial}{\partial{r}}+\frac{\beta+k_\sigma}{\beta r}\right)f^\sigma(r,\eta)&=&-(-1)^\sigma i^{l_\sigma-l_\sigma'}\left(\partial_\eta-\frac{1}{\eta}\left(\frac{3}{2}+\frac{i m\alpha}{\eta}\right)\right)g^\sigma(r,\eta)\nonumber\\
\left(\frac{\partial}{\partial{r}}+\frac{\beta-k_\sigma}{\beta r}\right)g^\sigma(r,\eta)&=&(-1)^\sigma i^{l_\sigma-l_\sigma'}\left(\partial_\eta-\frac{1}{\eta}\left(\frac{3}{2}-\frac{im\alpha}{\eta}\right)\right)f^\sigma(r,\eta)
\end{eqnarray}
Squaring the above equations leads to two second order differential equations,
\begin{eqnarray}
\label{frtau1'}
\left(\frac{\partial^2}{\partial r^2}+\frac{2}{r}\frac{\partial}{\partial{r}}-\frac{k_\sigma\left(k_\sigma+\beta\right)/\beta^2}{r^2}\right)f^\sigma(r,\eta)-\left(\frac{\partial^2}{\partial \eta^2}-\frac{3}{\eta}\frac{\partial}{\partial{\eta}}+\frac{\left(15/4-im\alpha+m^2\alpha^2\right)}{\eta^2}\right)f^\sigma(r,\eta)&=&0\nonumber\\
\left(\frac{\partial^2}{\partial r^2}+\frac{2}{r}\frac{\partial}{\partial{r}}-\frac{k_\sigma\left(k_\sigma-\beta\right)/\beta^2}{r^2}\right)g^\sigma(r,\eta)-\left(\frac{\partial^2}{\partial \eta^2}-\frac{3}{\eta}\frac{\partial}{\partial{\eta}}+\frac{\left(15/4+im\alpha+m^2\alpha^2\right)}{\eta^2}\right)g^\sigma(r,\eta)&=&0\nonumber\\
\end{eqnarray}
Separating now the variables, $f^{\sigma}(r,\eta)=T^{\sigma}_1(\eta)R^{\sigma}_1(r)$ (no sum on $\sigma$) we have
\begin{eqnarray}
\label{frtau2}
\frac{d^2 R_1^{\sigma}(r)}{dr^2}+\frac{2}{r}\frac{dR_1^{\sigma}(r)}{dr}+\left(\lambda^2-\frac{k_\sigma\left(k_\sigma+\beta\right)/\beta^2}{r^2}\right)R_1^{\sigma}(r)&=&0,\nonumber\\
\frac{d^2 T_1^{\sigma}(\eta)}{d\eta^2}-\frac{3}{\eta}\frac{dT_1^{\sigma}(\eta)}{d\eta}+\left(\lambda^2+\frac{15/4+m^2\alpha^2-im\alpha}{\eta^2}\right)T_1^{\sigma}(\eta)&=&0,
\end{eqnarray}
where $\lambda$ is a separation constant.
Likewise we take $g^{\sigma}(r,\eta)=T^{\sigma}_2(\eta)R^{\sigma}_2(r)$, to obtain
\begin{eqnarray}
\label{frtau3}
\frac{d^2 {R}_2^{\sigma}(r)}{dr^2}+\frac{2}{r}\frac{d{R}_2^{\sigma}(r)}{dr}+\left(\lambda^2-\frac{k_\sigma\left(k_\sigma-\beta\right)/\beta^2}{r^2}\right){R}_2^{\sigma}(r)&=&0\nonumber\\
\frac{d^2{T}_2^{\sigma}(\eta)}{d\eta^2}-\frac{3}{\eta}\frac{d{T}_2^{\sigma}(\eta)}{d\eta}+\left(\lambda^2+\frac{15/4+m^2\alpha^2+im\alpha}{\eta^2}\right){T_2}^{\sigma}(\eta)&=&0
\end{eqnarray}
The solutions to the spatial parts of \ref{frtau2}
and \ref{frtau3} are given by the spherical Bessel's functions

\begin{eqnarray}
\label{rad_sol1}
R^{\sigma}_1(r)\sim  \frac{1}{\sqrt r}{J_{\nu_\sigma}(\lambda r)},\qquad R^{\sigma}_2(r)\sim \frac{1}{\sqrt r }{J_{{\nu_\sigma}+(-1)^\sigma}(\lambda r)}
\end{eqnarray}
where 
$$\nu_{\sigma}=k_\sigma/\beta+1/2$$
 and $k_{\sigma}$ is given by \ref{opertaor1'}.

We take the temporal parts of  \ref{frtau2}
and \ref{frtau3} as the Hankel functions of the first kind
\begin{eqnarray}
\label{tempo_sol1}
T^{{\sigma}}_1(\eta)\sim \eta^2H^{(1)}_{1/2-im\alpha}(\lambda|\eta|),\qquad 
T^{{\sigma}}_2(\eta)\sim \eta^2 H^{(1)}_{-1/2-im\alpha}(\lambda|\eta|)
\end{eqnarray} 
Thus, for the spinor $\Psi_1$ in \ref{Psi1}, we may take
\begin{eqnarray}
\label{psi3}
\Psi_1=\sqrt{\frac{\pi}{2\lambda r}}{\eta^2}
   H^{(1)}_{1/2-im\alpha}(\lambda|\eta|){J_{\nu_\sigma}}(\lambda r)\Omega_{jl_\sigma m}
\end{eqnarray}
Using then \ref{dirac3}, \ref{Omega} we also have
\begin{eqnarray}
\label{psi4}
\Psi_2=-i(\hat{r}\cdot{\vec{\sigma}})\sqrt{\frac{\pi}{2\lambda r}}{\eta^2}
 H^{(1)}_{-1/2-im\alpha}(\lambda|\eta|){J_{\nu_\sigma+(-1)^\sigma}}(\lambda r)\Omega_{jl_\sigma m}
\end{eqnarray}
Recalling now the asymptotic form of the Hankel functions~\cite{Abra64} in the asymptotic past, $\eta \to -\infty$,
\be
H^{(1)}_{1/2-im\alpha}(\lambda|\eta|)\approx \sqrt{\frac{2}{\pi \lambda|\eta|}}e^{i(-\lambda\eta-\pi/4)}e^{-m\pi\alpha/2}
\label{Hankel}
\ee
 we identify the two positive frequency solutions (corresponding to $\sigma=0,1$ for $\lambda >0$) as
\begin{eqnarray}
\label{psi7}
\Psi^{(+)}_{\sigma j l m}&=&\sqrt{\frac{\pi}{2\lambda r}}{N_{\sigma}}\left( {\begin{array}{cc}
  \eta^2 H^{(1)}_{1/2-im\alpha}(\lambda|\eta|){J_{\nu_\sigma}}(\lambda r)\Omega_{jl_\sigma m}\\
   -i(\hat{r}\cdot{\vec{\sigma}}) \eta^2 H^{(1)}_{-1/2-im\alpha}(\lambda|\eta|){J_{\nu_\sigma+(-1)^\sigma}}(\lambda r)\Omega_{jl_\sigma m} \\
  \end{array} } \right),
\end{eqnarray}

The normalisation constant $N_{\sigma}$ can be determined via the relation,
\begin{eqnarray}
\label{norm}
\frac{\beta^2\alpha^{3}}{|\eta|^3}\int r^2 \sin \theta dr d\theta d\phi\, \left(\Psi^{(+)}_{\sigma{j}{l}{m}}\right)^{\dagger}{\Psi^{(+)}_{\sigma'{j'}{l'}{m'}}}=\delta_{\sigma \sigma'}\delta_{j j'}\delta_{l l'}\delta_{m m'}
\end{eqnarray}
Using \ref{psi7}, the left hand side of the above equation  becomes
\begin{eqnarray}
\frac{|N_\sigma|^2 \beta^2 |\eta|\pi}{2\lambda\alpha^{-3}}\left(|H^{(1)}_{1/2-im\alpha}(\lambda|\eta|)|^2+|H^{(1)}_{-1/2-im\alpha}(\lambda|\eta|)|^2\right)\int_0^\infty r drJ_{\nu_\sigma}(\lambda{r})J_{\nu_\sigma'}(\lambda{r})\int_0^{\pi}\int_0^{2\pi} d\phi\,d\theta \sin\theta\,\Omega_{jl m}^{\dagger}\Omega_{j' l'{m}'} \nonumber\\
\end{eqnarray}
Using now the integrals,
\begin{eqnarray}
\label{thth-phph}
\int_0^{\pi}\int_0^{2\pi}d\theta d\phi \,\sin\theta\,\Omega_{jl m}^{\dagger}\Omega_{j' l' m'} =\delta_{j j'}\delta_{ll'}\delta_{mm'},\qquad \int_0^\infty r drJ_{\nu_\sigma}(\lambda{r})J_{\nu_\sigma'}(\lambda{r})=\frac{1}{\lambda}{\delta_{\sigma\sigma'}}
\end{eqnarray}
and the asymptotic form of the  Hankel function, \ref{Hankel}, we have 
\begin{eqnarray}
\label{normfinal}
|N_\sigma|^2=\frac{\lambda^3}{2\beta^2\alpha^{3} }e^{\pi{m}\alpha}
\end{eqnarray}
Note that the  normalisation is independent of  $\sigma$. \\

The negative frequency modes are found via the charge conjugation : $\Psi^{(-)}_{\sigma j l m}=i \gamma^{(2)}{\Psi^{(+)^*}_{\sigma j l m}}$,
\begin{eqnarray}
\label{psi8}
\Psi^{(-)}_{\sigma j l m}(\eta,r,\theta,\phi)&=&\sqrt{\frac{\pi}{2\lambda r}}M_{\sigma}\left( {\begin{array}{cc}
  -i(-1)^\sigma(\hat{r}\cdot{\vec{\sigma}})\eta^2 H^{(2)}_{-1/2+im\alpha}(\lambda|\eta|){J_{\nu_\sigma+(-1)^\sigma}}(\lambda r)(i\sigma^2\Omega_{jl_\sigma m}^*)\\
   \eta^2H^{(2)}_{1/2+im\alpha}(\lambda|\eta|){J_{\nu_\sigma}}(\lambda r)(i\sigma^2\Omega_{jl_\sigma m}^*), \\
  \end{array} } \right)
\end{eqnarray}
where we have used $(H^{(1)}_{\alpha}(x))^*=H^{(2)}_{\alpha^{\star}}(x)$, for real $x$. Using now the well known properties of the spin harmonics,
\begin{eqnarray}
\label{psi11}
(i\sigma^2)\Omega_{l+1/2,l,m}^*=(-1)^{m+1/2}\Omega_{l+1/2,l,-m},\quad (i\sigma^2)\Omega_{l-1/2,l,m}^*=(-1)^{m+3/2}\Omega_{l-1/2,l,-m},\quad  
\Omega_{j l,-m}=(-1)^{m+1/2}\Omega_{jlm} \nonumber\\
\end{eqnarray}
\ref{psi8} simplifies to
\begin{eqnarray}
\label{psi9}
\Psi^{(-)}_{\sigma j l m}(\eta,r,\theta,\phi)&=&\sqrt{\frac{\pi}{2\lambda r}}M_{\sigma}\left( {\begin{array}{cc}
  i(-1)^\sigma(\hat{r}\cdot{\vec{\sigma}})\eta^2 H^{(2)}_{-1/2+im\alpha}(\lambda|\eta|){J_{\nu_\sigma+(-1)^\sigma}}(\lambda r)\Omega_{jl_\sigma m}\\
   \eta^2H^{(2)}_{1/2+im\alpha}(\lambda|\eta|){J_{\nu_\sigma}}(\lambda r)\Omega_{jl_\sigma m} \\
  \end{array} } \right)
\end{eqnarray}
where
\begin{eqnarray}
|M_{\sigma}|^2=\frac{\lambda^3 }{2\beta^2 \alpha^{3}}e^{-\pi{m}\alpha}
\end{eqnarray}
%

\section{$\langle 0| \overline{\Psi} \Psi | 0\rangle$ for pure de Sitter spacetime}\label{B}
The fermionic vacuum condensate for the pure dS spacetime is obtained by setting $\beta=1$ in \ref{conds4'}, 
\begin{eqnarray}
\label{dsgms-1}
\langle{0}|\overline{\Psi}\Psi|{0}\rangle_{\text{dS}}
=\frac{\eta }{\pi^2 \alpha^{3} r}\int_0^\infty {dy}y\;e^{y\left(1-r^2/\eta^2\right)}\text{Im}[K_{1/2-im\alpha}(y)]%
\sum_{j=1/2}^{\infty}(j+1/2)\left(I_{\left(j+1/2\right)+1/2}(yr^2/\eta^2)+I_{\left(j+1/2\right)-1/2}(yr^2/\eta^2)\right)\nonumber\\
\end{eqnarray}
Note that $(j+1/2)$ appearing above takes values $1,2,3, \dots$. Using then the formula~\cite{EMOT}
\begin{eqnarray}
\label{Modified}
\sum_{k=1}^{\infty}kI_{k+\nu}(z)=\frac{e^z}{2}\int_{0}^z ds e^{-s}I_\nu(s),
\end{eqnarray}
we obtain after some algebra
\begin{eqnarray}
\label{dsgms-2}
{\langle{0}|\overline{\Psi}\Psi|{0}\rangle_{\text{dS}}}&=&{\frac{\eta }{2\pi^2 \alpha^{3} r}\int_0^\infty {dy}\;y\;e^y\text{Im}[K_{1/2-im\alpha}(y)]}\,{\int_{0}^{{yr^2/\eta^2}}ds e^{-s}\left(I_{1/2}(s)+I_{-1/2}(s)\right)}
\end{eqnarray}
Using now
 ${I_{1/2}(s)=\sqrt{\frac{2}{\pi{s}}}\cosh s\;\text{and}\;I_{-1/2}(s)=\sqrt{\frac{2}{\pi{s}}}\sinh s}$ (e.g.~\cite{EMOT}), \ref{dsgms-2} simplifies  to
\begin{eqnarray}
\label{dsgms-3}
{\langle{0}|\overline{\Psi}\Psi|{0}\rangle_{\text{dS}}}&=&{\frac{8}{\left(2\pi\right)^{5/2} \alpha^3}\int_0^\infty {dy}\;y^{3/2}\;e^y\;\text{Im}[K_{1/2-im\alpha}(y)]}
\end{eqnarray}
Expectedly, the above integral is divergent. In order to regularise it, we introduce an exponential cut-off by making the replacement, $e^y \to e^{(1-\epsilon)y}$  ($\epsilon >0$), so that 

\begin{eqnarray}
\label{dsgms-5}
{\langle{0}|\overline{\Psi}\Psi|{0}\rangle_{\text{dS}}^{\left(\epsilon\right)
}}&=&{\frac{1}{\pi  \alpha^{3} \sinh(\pi m\alpha)}\partial_\epsilon^2{}_2F_{1}\left(i m\alpha,-i m\alpha+1;1;1-\frac{\epsilon}{2}\right)}
\end{eqnarray}
where we also have used the integral relationship given in \cite{Grad}.
The above can be simplified to (e.g.~\cite{Abra64}),
\begin{eqnarray}
\label{final_1'}
\langle{0}|\overline{\Psi}\Psi|{0}\rangle_{\text{dS}}^{\left(\epsilon\right)
}=-\frac{m }{4\pi^2\alpha^{2}}\bigg\lbrace \frac{2}{\epsilon}+\left(1+m^2\alpha^2\right)\ln \frac{\epsilon}{2}+2\left(1+m^2\alpha^2\right)\left[ \text{Re}\,\psi(im\alpha)-\ln\left(m\alpha\right)+\gamma-\frac34\right]+\mathcal{O}\left(\epsilon\right)\bigg\rbrace,
\end{eqnarray}
where $\gamma$ is Euler's constant and $\psi$ is the digamma function. Note  that the above expression is independent of the spacetime,  owing to the maximal symmetry of the pure de Sitter background. Also  as expected, it is the same as that of~\cite{BezerradeMello:2010ci}, found in the context of the de Sitter cosmic string background using a cylindrical coordinate. 

Note also that owing to the $\epsilon$-derivative in \ref{dsgms-5}, \ref{final_1'} is unique only up to some $\epsilon$-independent additive constant.
This ambiguity can be tackled by  imposing the physical condition that in the heavy mass limit,  $m\to \infty$, the condensate must vanish~\cite{BezerradeMello:2010ci}. Since the vacuum  expectation value of the trace of the fermionic energy-momentum tensor is given by $m\langle \overline{\Psi}\Psi \rangle $, it is clear that the  divergences of \ref{final_1'} can be absorbed by the renormalisation of the cosmological constant. After renormalising, and choosing the aforementioned $\epsilon$-independent additive constant appropriately, we have the final expression  of \ref{final_2}. It is easy to check by expanding the digamma function  of \ref{final_2} that in the large mass limit, it indeed vanishes as ${\cal O}(m^{-1})$.

\section{Energy-momentum tensor in pure de Sitter background}\label{C}

In this Appendix we shall very briefly sketch the derivation of the vacuum expectation value of the energy-momentum tensor in the pure de Sitter spacetime ($\beta=1$ in \ref{metric2}). For similar derivation in the context of the de Sitter cosmic strings, we refer our reader to~\cite{BezerradeMello:2010ci}. 

Now, setting $\beta=1$ in \ref{energy-density3} or \ref{rr-emt1}, and using the identity \ref{Modified}, we obtain 
\begin{eqnarray}
\label{ds-emt-tt}
\langle{0}|{T_0{}^0}|{0}\rangle_{\text{dS}}&=&\frac{\eta }{4\pi^2 \alpha^{4}{r}}\sum_{j=1/2}^{\infty}\left(2j+1\right)\int_0^\infty{dy}\;y\;e^{y\left(1-r^2/\eta^2\right)}\Bigg(I_{{j+1/2}+1/2}(yr^2/\eta^2)+I_{{j+1/2}-1/2}(yr^2/\eta^2)\Bigg)\text{Re}[K_{1/2-im\alpha}(y)]\nonumber\\
&=&{\frac{4}{\left(2\pi\right)^{5/2}\alpha^{4}}\int_0^\infty{dy}\;y^{3/2} e^{y}\text{Re}[K_{1/2-im\alpha}(y)]} = \langle{0}|{T_r{}^r}|{0}\rangle_{\text{dS}}
\end{eqnarray}
Likewise, by setting $\beta=1$ in \ref{11-theta-theta-3'}, we have
\begin{eqnarray}
\label{11-theta-theta-3}
{\langle{0}|{T_\theta{}^{\theta}}|{0}\rangle_{\rm dS}}&=&{\langle{0}|{T^\phi_\phi}|{0}\rangle_{\rm dS}=\frac{4}{\left(2\pi\right)^{5/2}\alpha^{4}}\int_0^\infty{dy}\;y\;e^{y\left(1-r^2/\eta^2\right)}\text{Re}\left[K_{1/2-im\alpha}(y)\right]}\left(y\partial_y-yr^2/\eta^2+1/2\right)\left(e^{yr^2/\eta^2} y^{1/2}\right)\nonumber\\
&=&{\frac{4}{\left(2\pi\right)^{5/2}\alpha^{4}}\int_0^\infty{dy}{y}^{3/2}e^{y}\text{Re}\left[K_{1/2-im\alpha}(y)\right]}
\end{eqnarray}

As expected, the above expressions are  the same as that 
of~\cite{BezerradeMello:2010ci}, found in the context of the de Sitter cosmic string. 
Note that all the components of $\langle{0}|{T_\mu{}^\nu}|{0}\rangle_{\rm dS}$ are equal and independent of the spacetime coordinates, as expected from the maximal symmetry of the de Sitter.   The components of the energy-momentum tensor can  thus be put in a  compact form,
\begin{eqnarray}
\label{emt-ds-pure}
\langle{0}|{T_\mu{}^\nu}|{0}\rangle_{\rm dS}=\frac{4\delta_{\mu}{}^\nu}{\left(2\pi\right)^{5/2}\alpha^{4}}\int_0^\infty{dy}\;{y}^{3/2}e^{y}\text{Re}\left[K_{1/2-im\alpha}(y)\right]
\end{eqnarray} 
 The  renormalised expression can be found after using a cosmological constant counterterm~\cite{BezerradeMello:2010ci}, 
\begin{eqnarray}
\label{emt-ds-ren-pure}
\langle{0}|{T_\mu^\nu}|{0}\rangle_{\rm Ren.,~dS}=\frac{\delta_\mu{}^\nu}{8\pi^2 \alpha^{4}}\left[m^2\alpha^2\left(1+m^2\alpha^2\right)\left[\text{ln}\left(m\alpha\right)-\text{Re}\left(i m\alpha\right)\right]+\frac{m^2\alpha^{2}}{12}+\frac{11}{120}\right]
\end{eqnarray}
%


\end{document}